\documentclass[12pt]{article}
\pdfoutput=1
\usepackage{graphics}
\usepackage[dvips]{graphicx}
\usepackage{mathrsfs}
\usepackage{amssymb}
\usepackage{verbatim}
\usepackage{float}
\newcommand{\be}{\begin{equation}}
\newcommand{\ee}{\end{equation}}
\newcommand{\bea}{\begin{eqnarray}}
\newcommand{\eea}{\end{eqnarray}}

\def\4vol{{\int d^4x \sqrt{-g}}}

\def\beq{\begin{equation}}
\def\eeq{\end{equation}}
\def\bitem{\begin{itemize}}
\def\eitem{\end{itemize}}

\newcommand{\nc}{\newcommand}

\nc{\nt}{\tilde{N}}
\nc{\ra}{\rightarrow}
\nc{\lsim}{\begin{array}{c}\,\sim\vspace{-21pt}\\< \end{array}}
\nc{\gsim}{\begin{array}{c}\sim\vspace{-21pt}\\> \end{array}}
\nc{\tnt}{\tilde{N}}
\nc{\tst}{\tilde{t}}

\nc{\LL}{L}
\nc{\vv}{\tilde{v}}

\title{
\vspace*{5mm} \Large\textbf{Warped Universal Extra Dimensions}
\vspace*{1.0cm}
\author{\textbf{An\'ibal D.~Medina~$^a$ and Eduardo Pont\'on~$^b$}\\\\
\normalsize\emph{$^a$Department of Physics, University of California,
One Shields Ave.  Davis, CA 95616, USA}\\
\normalsize\emph{$^b$Department of Physics, Columbia University,
538 W. 120th St, New York, NY 10027, USA}}}
\date{\today}
\begin{document}
\setcounter{page}{0}
\maketitle
\begin{abstract}

We consider a 5D warped scenario with a KK-parity symmetry, where the
non-trivial warping arises from the dynamics that stabilizes the size
of the extra dimension.  Generically, the lightest Kaluza-Klein (KK)
particle is the first excitation of the radion field, while the
next-to-lightest Kaluza-Klein particle is either the first excitation
of the (RH) top quark or the first KK-parity odd Higgs.  All these
masses are expected to be of order the electroweak scale.  We present
simple analytical expressions for the masses and wavefunctions of the
lowest lying KK modes, and derive the Feynman rules necessary for
phenomenological applications.  The framework allows to interpolate
between a strongly warped scenario \'a la Randall-Sundrum (RS), and a
weakly warped scenario that shares properties of both RS and Universal
Extra Dimensions models.

\end{abstract}

\thispagestyle{empty}
\newpage
\setcounter{page}{1}

\section{Introduction}
\label{Intro}

Warped extra dimensional scenarios provide an appealing solution to
the hierarchy problem~\cite{Randall:1999ee}.  In addition, when the SM
gauge bosons~\cite{Davoudiasl:1999tf,Pomarol:1999ad} and
fermions~\cite{Grossman:1999ra} arise from higher-dimensional fields
one gains a rather interesting understanding of the SM fermion
hierarchies and CKM mixing angles, based on the localization of these
modes along the extra dimension~\cite{Gherghetta:2000qt}: light
fermion fields are localized near the ``UV brane'' while heavier
fermions, most notably the top quark, are localized closer to the ``IR
brane''.  At the same time dangerous flavor-changing neutral current
effects are suppressed, and can be roughly consistent with low-energy
constraints even when the new physics is at the TeV scale.  The
associated spectrum of Kaluza-Klein (KK) modes provides exciting
signals at the LHC.

It was recognized early on~\cite{Sundrum:1998ns,ArkaniHamed:1998kx}
that a complete solution to the hierarchy problem requires the
specification of dynamics that stabilizes the size of the extra
dimensions.  In the Randall-Sundrum (RS)
scenario~\cite{Randall:1999ee}, this requires that the interbrane
separation be stabilized at a factor of $\sim 35$ in units of the
inverse curvature scale.  This can be achieved either at tree-level,
by considering for example a SM singlet scalar that acquires a
non-trivial vacuum expectation value (VEV)~\cite{Goldberger:1999uk},
or at the quantum level (see
e.g.~\cite{Garriga:2000jb,Goldberger:2000dv,Garriga:2002vf,Bai:2008gm,Maru:2010ap,Davoudiasl:2010pk}).
An important consequence is the presence of a 4D scalar mode, called
the ``radion'', that corresponds to the fluctuations in the interbrane
distance.  The radion mode is expected to be parametrically lighter
than the KK states, and therefore potentially important in the
phenomenology of these scenarios.  The generic couplings of the radion
to SM fields were worked out in
Refs.~\cite{Goldberger:1999un,Csaki:2000zn,Rizzo:2002pq,Csaki:2007ns}.
Except for a possible mixing term with the Higgs
field~\cite{Csaki:1999mp,Giudice:2000av,Dominici:2002jv}, these
interactions are non-renormalizable and suppressed by a ``radion decay
constant'', $\Lambda_{r}$.

In this paper we propose stabilizing the size of the
extra-dimension, parameterized by $y \in [-L,L]$, in such a way
that the non-trivial warping arises from the backreaction of a
Goldberger-Wise (GW) scalar, with the geometry being symmetric
about $y=0$.  As a result, the boundaries of the space, at $y =
\pm L$, are IR boundaries were the warp factor is smallest, while
the maximum of the warp factor is achieved at $y = 0$.  Thus,
although no UV brane is put in by hand, a dynamical ``UV brane''
is generated in the middle of the extra dimension.  In fact, in a
certain limit this UV brane becomes infinitely thin, and the
framework reduces to two copies of the RS
scenario~\cite{Randall:1999ee} glued together at the UV brane, as
considered in~\cite{Agashe:2007jb}.  However, generically, the
dynamical UV brane is ``fat'', and the geometry is not simply
AdS$_{5}$.

One important consequence of the setup, that goes beyond the most
studied implementations of the RS proposal (for a review, see
e.g.~\cite{Davoudiasl:2009cd}), is the existence of a discrete $Z_{2}$
symmetry under which all the even-level KK states (which include the
zero-modes) are $Z_{2}$-even, while the odd-level modes are
$Z_{2}$-odd (i.e.~a KK parity similar to the one present in Universal
Extra Dimensional scenarios or
UED's~\cite{Appelquist:2000nn,Dobrescu:2004zi}).  As a result, the
lightest KK-parity odd particle (the LKP) is stable, and a potential
dark matter (DM) candidate (for an alternative proposal,
see~\cite{McDonald:2009md}).  However, unlike in UED
scenarios~\cite{Cheng:2002iz,Ponton:2005kx}, the KK states are
\textit{not necessarily} highly degenerate, with the degeneracy broken
by loop-level effects, but instead larger mass splittings can be
expected (although there are some interesting degeneracies, related to
localization effects).  A similar idea of a warped framework with a KK
parity was explored in~\cite{Agashe:2007jb}, where the viability of
the first KK excitation of the $Z$ gauge boson as a DM candidate was
investigated.  Here, we point out that the LKP is generically expected
to be the first KK excitation of the radion field (more precisely of
the radion/GW-scalar system).  We show that its mass is highly
degenerate with the radion mass, and that its interactions are also
controlled by $\Lambda_{r}$.  Then, the question of the viability of
the first KK radion as a DM candidate naturally arises, and answering
it becomes essential in such warped frameworks with a KK parity.  The
point is that, although it is expected that the LKP has a mass of
order the EW scale, its interactions may not be sufficiently strong to
ensure the annihilation of KK radions during the early history of the
universe to levels below the ``overclosure limit''.  The WIMP miracle
is not necessarily automatic, and the danger of overproducing the
LKP's becomes a pressing issue.  Fortunately, there are a number of
scenarios one can envision where the KK radion relic density can be
suppressed and explain the observed DM density.  We explore these in
detail in the companion paper~\cite{Medina:2011qc}.

In this work we lay out the formalism necessary to analyze the
properties of the LKP, which includes also a discussion of the KK
excitations of fermion and gauge fields.  Since we consider a rather
general gravitational background, it is difficult to find closed
solutions for the KK wavefunctions.  However, we
obtain reasonably accurate analytical approximation for the lightest
states in the theory, which include the radion, the KK radion and
possibly the first KK excitation of the (RH) top.  These become more
accurate in the limit that the UV brane is fat, which corresponds to
an interesting deformation of the RS proposal.  We also point out that
fermion zero-mode localization is naturally achieved by coupling the 5D
fermions to the GW scalar that induces the warping and stabilizes the
size of the extra dimension.  Thus, no ad-hoc 5D fermion masses need
to be introduced.  Regarding the EWSB sector, it is shown that it is
naturally a two-Higgs doublet model (THDM), with the second doublet
being a KK-parity odd excitation of the SM Higgs.

Our formalism also allows us to interpolate between a strongly warped
limit (\'a la RS) and a limit where the curvature is small, which is
similar to the UED assumption.  An interesting intermediate case
arises when the curvature is small compared to the 5D Planck mass, but
not exactly zero.  In this case, the hierarchy between the Planck and
the electroweak (EW) scales arises in part from a moderate warp
factor, and in part from the smallness of the curvature scale.  Such a
scenario can share phenomenological properties of both RS and UED
scenarios.  It can also give rise to KK radion DM as a non-thermal
relic~\cite{Medina:2011qc}.  Our approach is somewhat different from
the recently studied ``soft-wall''
scenarios~\cite{Gubser:2000nd}--\cite{Cabrer:2010si}, but it could be
interesting to combine the two ideas to obtain a fully dynamical
warped compactification.

The outline of this work is as follows: in Section~\ref{Basic} we give
several examples that illustrate the stabilization of the extra
dimension with an automatic KK-parity symmetry.  We explore various
limits, imposing a minimal number of phenomenological constraints, and
define two benchmark scenarios to be used as reference.  In
Section~\ref{RadionOddScalar}, we discuss the KK decomposition of the
radion/GW-scalar system, thus defining the radion decay constant
$\Lambda_{r}$.  We also give simple analytical expressions for the
mass and wavefunction profiles of both the radion and its first KK
excitation (the LKP).  Section~\ref{sec:Fermion} is devoted to a study
of fermions in the $Z_{2}$ symmetric backgrounds.  We point out that
in the fat UV brane limit it is possible to characterize the
localization of the zero-mode via a dimensionless $c$-parameter that
is analogous to the treatment in the pure AdS$_{5}$ background,
although there are important differences.  We also give approximate
analytical expressions for the mass and wavefunction of the first
fermion KK resonance in cases where it is parametrically lighter than
the KK scale (hence the likely next-to-lightest KK-parity odd
particle, or NLKP).  Section~\ref{sec:Higgs} is devoted to bulk
scalars as applied to the Higgs field.  We also present the Higgs
Yukawa interactions with bulk fermions.  Gauge fields are briefly
discussed in Section~\ref{sec:gauge}.  In preparation to the DM
analysis to be presented in~\cite{Medina:2011qc}, we work out in
Section~\ref{sec:Rcouplings} the Feynman rules for the interactions
between the radion/KK-radion and the KK fermions, gauge bosons and the
Higgs doublets.  We summarize and conclude in
Section~\ref{sec:conclusions}.

\section{General Setup}
\label{Basic}

In this section we describe the stabilization mechanism that leads to
a non-trivial $Z_{2}$-symmetric warping.  We illustrate this by
choosing scalar potentials that allow for closed analytical solutions
that take into account the scalar backreaction on the metric, but
emphasize that the properties we are interested in do not depend on
any given particular choice of potential.  We discuss an initial set
of phenomenological constraints, and define two benchmark scenarios,
one with strong warping and one where the curvature is small compared
to the 5D Planck scale.

\subsection{Backgrounds with $Z_{2}$ Symmetry}
\label{Background}

We start with a 5D real scalar $\Phi$ minimally coupled to gravity
according to
\bea
S = \int_{M} \! d^5 x \,
\sqrt{g} \left[ -\frac{1}{2} M^{3}_{5} \, \mathcal{R}_{5} + \frac{1}{2} \nabla_M \Phi
\nabla^M \Phi  -V(\Phi)\right]
+ \int_{\partial M} \! d^4x \, \sqrt{g_{\rm ind}} \, {\cal L}_{4}(\Phi)~,
\label{GravityScalarAction}
\eea
where $M_{5}$ is the (reduced) 5D Planck mass, $\mathcal{R}_{5}$ is
the 5D Ricci scalar, and the last term allows for operators localized
on the boundary of the fifth dimension (which will be specified in the
next subsection).  We assume that the fifth dimension is compactified
to an interval, parameterized by $y \in [-L,L]$.  Our interest is in
dynamics that stabilizes the size of the extra dimensions, $2L$, while
leading to a background that is symmetric about $y = 0$, so that a
geometric $Z_{2}$ symmetry\footnote{As established in
\cite{Bai:2009ij}, Chern-Simons terms do not break the KK-parity
symmetry, so that the questions raised in~\cite{Hill:2007zv} for
little Higgs scenarios with T-parity do not apply.} is present, as
in~\cite{Agashe:2007jb}.  However, unlike Ref.~\cite{Agashe:2007jb} we
do not have a central brane at $y=0$, although as we will see, in
certain limits, such a brane arises from the scalar dynamics.

We are interested in solving for the coupled gravity/scalar system,
taking into account the backreaction of a scalar VEV on the geometry,
since this results in the stabilization of the extra dimension \`a la
Goldberger-Wise~\cite{Goldberger:1999uk}.  We restrict to backgrounds
exhibiting 4D Lorentz symmetry, which can always be put in the form
\bea
ds^2 &=& e^{-2A(y)} \eta_{\mu\nu} dx^\mu dx^\nu - dy^2~,
\label{metric}
\eea
where $y$ is the proper distance.  The weak energy condition
implies that $A''(y) \geq 0$~\cite{Freedman:1999gp} so that
$e^{-A(y)}$ must be a convex function and the two boundaries at $y
= \pm L$ are IR branes in the RS terminology.  As shown in
Refs.~\cite{Brandhuber:1999hb,DeWolfe:1999cp}, Einstein's
equations, together with the scalar equation of motion, can be
solved exactly for the class of scalar potentials given by
\bea
V(\Phi) = \frac{1}{8} \left(\frac{\partial W(\Phi)}{\partial \Phi} \right)^2 -
\frac{1}{6M^{3}_{5}} W(\Phi)^2~,
\label{PotFromSuperpot}
\eea
where $W(\Phi)$ is an arbitrary ``superpotential'' ($W$ has mass
dimension $4$, while $\Phi$ has mass dimension $3/2$ in 5D).  For the
present purpose, the important point is that the scalar background,
$\phi(y) \equiv \langle \Phi(y) \rangle$, can be obtained by solving
the \textit{first order} differential
equation~\footnote{Eqs.~(\ref{phiEOM}) and (\ref{AEOM}) guarantee that
the equations of motion for the gravity/scalar system are satisfied
when the potential is given by Eq.~(\ref{PotFromSuperpot});
see~\cite{DeWolfe:1999cp} for details.}
\bea
\phi'(y) &=& \frac{1}{2} \frac{\partial W(\phi)}{\partial \phi}~,
\label{phiEOM}
\eea
and that once the solution $\phi(y)$ is found, $A'(y)$ is simply given
by
\bea
A'(y)  &=& \frac{1}{6M^{3}_{5}} W(\phi(y))~,
\label{AEOM}
\eea
which can be integrated immediately to yield
$A(y)$.\footnote{Alternatively, given $A(y)$ one can combine
Eqs.~(\ref{phiEOM}) and (\ref{AEOM}) using $\partial_{\phi} =
(1/\phi') \partial_{y}$ to obtain $\phi' = \pm \sqrt{12M^{3}_{5}
A''(y)}$, which can be integrated to yield $\phi(y)$, up to a sign.
If $\phi(y)$ can be inverted to obtain $y = y(\phi)$, the
superpotential is given by $W(\phi) = 6M^{3}_{5} A'(y(\phi))$.}
Eq.~(\ref{phiEOM}) shows that $\phi'$ is even under $y \rightarrow
-y$, provided $\partial W/\partial\phi$ is even --hence $W(\phi)$ is
odd-- under $\phi \rightarrow -\phi$.  It then follows from
Eq.~(\ref{AEOM}) that $A'(y)$ is odd, and therefore $A(y)$ is even
under $y \rightarrow -y$, as we want.  Thus, imposing the $Z_{2}$
reflection symmetry allows us to focus on odd superpotentials.  We
also see that for the stabilizing scalar, $\Phi$, the discrete
symmetry should be a combination of the geometric reflection $y \to
-y$ and an internal parity symmetry: the action of the relevant
$Z_{2}$ on the GW scalar is $\Phi(x^\mu, y) \to - \Phi(x^\mu,-y)$.
The scalar potential derived from Eq.~(\ref{PotFromSuperpot}) is
clearly even under this $Z_{2}$, and the latter is respected by odd
``kink-like'' profiles for the VEV, $\phi(y)$.

For instance, the simplest ansatz for the superpotential is a linear
one:
\bea
W_{1}(\Phi) = 2m \phi_{0} \Phi~,
\label{LinearW}
\eea
where, for later convenience, we have parameterized the coefficient in
terms of a mass scale $m$ and a scalar ``VEV'' $\phi_{0}$ (with mass
dimension $3/2$).  Both Eqs.~(\ref{phiEOM}) and (\ref{AEOM}) can be
trivially integrated, and give
\bea
\phi(y) &=& \phi_{0} m y~,
\label{simplephi} \\[0.5em]
A(y) &=& \frac{\phi^{2}_{0} m^{2} }{6M^{3}_{5}} \, y^2~,
\label{simpleA}
\eea
where the condition $\phi(-y) = -\phi(y)$ sets the integration
constant to zero in Eq.~(\ref{simplephi}), and we choose $A(0)=0$ by
rescaling the 4D coordinates by the appropriate constant scale factor.
This simple example will turn out to capture the correct physics of a
large class of superpotentials, once the radion is stabilized and a
hierarchy is induced.

A richer example arises from a cubic superpotential:
\bea
W_{3}(\Phi) &=& \frac{2m}{\phi_{0}} \left( \phi_{0}^{2} \, \Phi - \frac{1}{3} \Phi^{3} \right)~,
\label{cubicW}
\eea
with $m$ an arbitrary mass scale that, without loss of generality, can
be chosen to be positive.  In this case, Eq.~(\ref{phiEOM}) reads
\bea
\phi'(y) &=& \frac{m}{\phi_{0}} \left[ \phi_{0}^{2} - \phi(y)^2 \right]~,
\label{kinkEqn}
\eea
which can be recognized to give the standard flat space ``kink''
solution~\footnote{The scalar potential, Eq.~(\ref{cubicV}), depends
only on $\phi_{0}^2$.  The sign of $\phi_{0}$ parameterizes the two
solutions related by $y \to -y$, which correspond to a
positive/negative VEV at $y=+\infty$.}
\bea
\phi(y) &=& \phi_{0} \tanh(my)~,
\label{cubicphi}
\eea
with a thickness of order $1/m$ and asymptotic VEV $\pm \phi_{0}$.
Following the above described procedure, one then finds
\bea
A(y) &=& \frac{\bar{k}}{4m} \left[\tanh^2 (my) + 4 \log\cosh (my) \right]~,
\label{cubicA}
\eea
where we defined the scale $\bar{k}$ by $\phi^{2}_{0} =
9\bar{k}M^{3}_{5}/(2m)$, and again we chose $A(0) = 0$.  Unlike the
example defined by the linear superpotential (\ref{LinearW}), this
model has the virtue that for $|y| \gg 1/m$ one has $A(y) \rightarrow
\bar{k}|y| + \textrm{const.}$, so that asymptotically one gets AdS$_{5}$
with curvature given by $\bar{k}$.  In fact, it is clear that this
corresponds to the limit where the domain-wall (\ref{cubicphi}) is
very narrow, i.e. $m \gg \bar{k}$, and models a UV brane at $y = 0$,
so that the RS solution is recovered.

We note that the potential that follows from the cubic superpotential
(\ref{cubicW}) is
\bea
V(\Phi) &=& \frac{9}{4} m \bar{k} M^{3}_{5} - m^{2} \left( 1 +
\frac{3\bar{k}}{m} \right) \Phi^{2} + \frac{4m^{2}}{9M^{3}_{5}}
\left( 1 + \frac{m}{4\bar{k}} \right) \Phi^{4} -
\frac{4m^{3}}{243\bar{k}M^{6}_{5}} \Phi^{6}~.
\label{cubicV}
\eea
From an effective field theory perspective, this is a scalar potential
subject to a parity symmetry $\Phi \rightarrow -\Phi$, and truncated
at sixth order in $\Phi$.  However, it is non-generic in two ways.
First, the constant term has a fixed value, which is necessary for the
ansatz of Eq.~(\ref{metric}) --with flat 4D sections-- to hold.  In
addition to this well-known Cosmological Constant Problem, we see that
the potential contains three operators that are controlled by only two
parameters, thus seemingly implying an additional fine-tuned relation
(at least in the absence of supersymmetry).  The two parameters can be
chosen as $\bar{k}/M_{5}$ and $m/M_{5}$ ($M_{5}$ simply sets the
overall scale, and is associated to the strength of the gravitational
interactions).  However, we stress that this additional relation plays
no essential role in our applications.  Indeed, we see that in the
thin domain-wall limit, $m \gg \bar{k}$, and using $|\phi(y)| <
\phi_{0}$, the ratio of the $\Phi^{6}$ operator to the $\Phi^{4}$
operator in Eq.~(\ref{cubicV}) is of order $\bar{k}/m \ll1$.  Hence,
the dynamics is controlled by the quadratic and quartic operators, the
last term in Eq.~(\ref{cubicV}) representing only a small
perturbation.  In the opposite limit, $m \rightarrow 0$, and for
$|\phi| \sim \phi_{0}$, the three operators are parametrically equally
important.  Nevertheless, in the central region, where $|\phi(y)| \ll
\phi_{0}$, the dynamics is actually controlled only by the quadratic
term, thus reducing precisely to the case of the linear superpotential
of Eq.~(\ref{LinearW}).  As we will see, this will be important in the
following and makes our results rather generic.  The point is that we
are using the special class of potentials that can be obtained from a
superpotential via Eq.~(\ref{PotFromSuperpot}) only to obtain simple,
closed solutions that fully take into account the backreaction of the
scalar on the geometry, but our physical results will not depend on
this simplifying assumption.

Our final example is based on the observation that the cubic
superpotential above corresponds to the first two terms in the Taylor
expansion of a sine:
\bea
W(\Phi) &=& \sqrt{2} m \phi^{2}_{0} \sin \left(\sqrt{2} \frac{\Phi}{\phi_{0}} \right)~.
\label{sineW}
\eea
This leads to a scalar profile
\bea
\phi(y) &=& \sqrt{2} \phi_{0} \tan^{-1}  \left[\tanh \left(\frac{m y}{\sqrt{2}} \right) \right]~,
\label{phivev}
\eea
that is also a domain wall of width of order $1/m$, centered at $y =
0$.~\footnote{The asymptotic VEV's are $\pm(\pi/2\sqrt{2})\,\phi_{0}
\approx \pm 1.11 \phi_{0}$.} Correspondingly,
\bea
A(y) &=& \frac{k}{\sqrt{2} m} \log \left[\cosh \left(\sqrt{2} m y \right) \right]~,
\label{ExplicitA}
\eea
where now $\phi_{0}$ and $k$ are related by $\phi^{2}_{0} =
6kM^{3}_{5}/(\sqrt{2}m)$, and we again made sure that the warp factor
is normalized to unity at $y=0$.  As in the second example, the
geometry is asymptotically AdS$_{5}$ with curvature $k$, i.e. one has
$A(y) \rightarrow k|y| + \textrm{const.}$ for $|y| \gg 1/m$.  Again,
the IR-UV-IR model of Ref.~\cite{Agashe:2007jb} arises in the limit $m
\gg k$.  Our comments above also apply to this example: although the
potential that follows from Eq.~(\ref{PotFromSuperpot}) with the
superpotential (\ref{sineW}) contains an infinite tower of operators,
only a few of them are actually relevant in determining the background
solution: the quadratic and quartic terms in the ``UV-brane'' limit,
$m \gg k$, and the quadratic term in the central region $y < 1/m$.
Nevertheless, the ``sine superpotential'' has the advantage that it
allows us to interpolate between narrow and wide domain-walls with a
simpler warp factor than for the cubic superpotential.  For
concreteness, in the numerical studies of the rest of this work we
will use Eqs.~(\ref{phivev}) and (\ref{ExplicitA}), with the
understanding that the results differ from the other (and similar)
examples only in minute details.

\subsection{Radion Stabilization}
\label{RadionStabilization}

We described in the previous subsection the \textit{bulk} solutions
that follow from a given superpotential.  However, one must ensure
that these solutions are actually allowed by the boundary conditions
at $y = \pm L$.  One can see immediately from Eqs.~(\ref{phiEOM}) and
(\ref{AEOM}) that this requires
\bea
\phi'(\pm L) &=& \frac{1}{2} \left. \frac{\partial W(\phi)}{\partial \phi} \right|_{y = \pm L}~,
\hspace{1cm}
A'(\pm L) ~=~ \left. \frac{1}{6M^{3}_{5}} W(\phi) \right|_{y = \pm L}~.
\label{BoundaryRestrictions}
\eea
On the other hand, the set of consistent boundary conditions can be
found by varying the action Eq.~(\ref{GravityScalarAction}) with
respect to both $\Phi$ and $A$, and requiring that the surface terms
obtained from the integrations by parts vanish.  Assuming that ${\cal
L}_{4}$ in Eq.~(\ref{GravityScalarAction}) does not involve $y$
derivatives, e.g. if it is a pure potential term, this leads
to~\cite{DeWolfe:1999cp}
\bea
\Phi'(\pm L) &=& - \frac{1}{2} \left. \frac{\partial {\cal L}_{4}}{\partial \Phi} \right|_{y = \pm L} ~,
\hspace{1cm}
A'(\pm L) ~=~ - \left. \frac{1}{6 M^{3}_{5}} \, {\cal L}_{4} \right|_{y = \pm L}~.
\label{BoundaryConditons}
\eea
Following Refs.~\cite{DeWolfe:1999cp,Csaki:2000zn} one can enforce
that the scalar field attain a given value $\bar{\phi}_{0}$ (in
absolute magnitude) at the boundaries by writing an even boundary
term~\footnote{One can also write $\Delta {\cal L}_{4} = -\gamma
\left( \Phi - \bar{\phi}_{0} \right)^{2}$ at $y = +L$, and $\Delta
{\cal L}_{4} = -\gamma \left( \Phi + \bar{\phi}_{0} \right)^{2}$ at $y
= -L$, so as to preserve the $\Phi(y) \rightarrow -\Phi(-y)$ symmetry.
In this case, the constant $\gamma$ has positive mass dimension and
perhaps it is more natural than in the case of
Eq.~(\ref{FixedBoundaryVEV}) for it to take a large value.  It also
has the virtue that it selects the ``domain-wall scalar profile'' over
a constant profile.  In any case, the most important (simplifying)
assumption is that the value of the scalar field be frozen on the
boundaries.}
\bea
\Delta {\cal L}_{4}(\Phi) &=& -\gamma \left( \Phi^2 - \bar{\phi}^2_{0} \right)^{2}~,
\label{FixedBoundaryVEV}
\eea
that does not contribute to Eqs.~(\ref{BoundaryConditons}) in the
limit $\gamma \rightarrow +\infty$.  Thus, we see that one can use
\bea
{\cal L}_{4}(\Phi) &=& -W (\bar{\phi}_{0}) - W'(\bar{\phi}_{0})(\Phi - \bar{\phi}_{0}) +  \Delta {\cal L}_{4}(\Phi)~,
\label{L4}
\eea
which leads to Eqs.~(\ref{BoundaryRestrictions}) [for $\gamma
\rightarrow +\infty$], as required.  In the limit of a fixed VEV,
$\bar{\phi}_{0}$, at the boundaries, we must have
\bea
\bar{\phi}_{0} &=& \phi(L) ~=~
\sqrt{2} \phi_{0} \tan^{-1}  \left[\tanh \left(\frac{m L}{\sqrt{2}} \right) \right]~,
\eea
where the second equality holds for the case of the ``sine''
superpotential, Eq.~(\ref{sineW}).  This equation fixes the size of
the extra dimension to be
\bea
m L &=& \sqrt{2} \tanh^{-1}  \left[\tan \left(\frac{\bar{\phi}_{0}}{\sqrt{2} \phi_{0}} \right) \right]~,
\label{mL}
\eea
and depends only on the ratio
$\bar{\phi}_{0}/\phi_{0}$.~\footnote{Notice that if
$\bar{\phi}_{0}/\phi_{0}$ is such that $|\tan
\left(\frac{\bar{\phi}_{0}}{\sqrt{2} \phi_{0}} \right)| > 1$ there are
no (real) solutions for $L$ from Eq.~(\ref{mL}).  This is simply the
statement that the brane VEV, $\bar{\phi}_{0}$, cannot be larger than
the asymptotic VEV of the bulk solution, Eq.~(\ref{phivev}), for the
configuration to be allowed.} Then Eq.~(\ref{ExplicitA}) gives
\bea
A(L) &=&
\frac{k}{\sqrt{2} m} \log \left[\cosh \left(\sqrt{2} m L \right) \right]~,
\label{AL}
\eea
which is parametrically large for $m \ll k$, i.e. away from the ``thin
UV brane'' limit, even when $mL \sim {\cal O}(1)$.  It is also large
in the limit that the extra dimension is large, $mL \gg 1$ and $kL \gg
1$, although this requires the brane and bulk VEV's to satisfy
$\bar{\phi}_{0}/(\sqrt{2} \phi_{0}) \approx \pi/4$, a tuned limit.

\subsection{From Large to Small Warping}
\label{Scales}

The model described above depends on three dimensionless ratios:
$k/m$, $\bar{\phi}_{0}/\phi_{0}$ and $k/M_{5}$.  The latter ratio
determines the degree of warping (from nearly flat when $k \ll M_{5}$,
to strongly warped when $k \sim M_{5}$).  The case of small warping
may require fine-tuning to make the 5D cosmological constant much
smaller than the fundamental scale $M_{5}$ (in addition to the
fine-tuning associated with the vanishing of the 4D CC).
Nevertheless, we will consider such a small warping limit since it
allows us to make the connection with the well-studied models of
Universal Extra Dimensions (UED)~\cite{Appelquist:2000nn}, which
assume that the 5D curvature is negligible, while still showing some
interesting differences when $k$ is small but non-vanishing.  In the
opposite limit of strong warping we can place a rough upper bound on
the size of the curvature from the requirement that there be some
modest hierarchy between $k$ and the cutoff of the theory,
$\Lambda_{5}$, so that neglecting higher-dimension operators in the
Einstein-Hilbert action is justified.  Based on NDA in
extra-dimensional theories~\cite{Chacko:1999hg} we expect that
$\Lambda^{3}_{5} \sim M^{3}_{5}/l_{5}$, with $l_{5} = 24\pi^{3}$, a 5D
loop factor.  This suggests that even for $k = M_{5}$, one has
$\Lambda_{5} \sim 10 k$, which is a large enough separation for the 5D
EFT to be useful.  Hence, whenever we need to be concrete, we will
take $k = M_{5}$ when discussing the strong warping limit, although
cases where $k$ is somewhat smaller can be considered.

We can further constrain the parameter space based on phenomenological
considerations.  First, one must reproduce the 4D Planck scale, $M_{P}
\approx 2.4 \times 10^{18}~{\rm GeV}$.  Second, as will be seen in the
following sections, the scale of KK resonances is set by $\tilde{k}
\equiv k \,e^{-A(L)}$.~\footnote{To be more precise, the KK scale is
set by the warped down curvature scale at the IR boundaries, $k_{\rm
eff} \,e^{-A(L)}$, where $k_{\rm eff} = A'(L) = s_{A} k$, and $s_{A}$
is a model-dependent number smaller than one.  See Eq.~(\ref{ApproxA})
below.} We will assume that $\tilde{k} \sim {\cal O}(1~{\rm TeV})$ to
open the window to test this scenarios at the LHC. These two
constraints determine $M_{5}$ and one of the three dimensionless
parameters mentioned above.  One can exchange one of the two remaining
parameters for the radion decay constant $\Lambda_{r}$ (to be
discussed in Section~\ref{RadionNormalization}), plus a free
dimensionless parameter that we will choose as $k/m$.  We elaborate
next on the first two relations, before discussing the properties of
the radion field.

\subsubsection{The 4D Planck Scale}
\label{PlanckMass}

We start by computing the effective 4D Planck scale from the
underlying Lagrangian parameters.  For instance, in the model defined
by Eq.~(\ref{sineW}) the 4D (reduced) Planck mass is
\bea
M_{P}^2 &=& M^{3}_{5} \int_{-L}^L e^{-2A(y)} dy
~=~ \frac{2M^3_{5}}{m} \int^{mL}_{0} \! dx \, \left[\cosh\left( \sqrt{2} x \right) \right]^{-\sqrt{2} k/m}~,
\label{MPlanck}
\eea
where the factor of $2$ arises from the $Z_{2}$ symmetry of the
background about $y=0$.  There are three important scales in the
problem: the curvature $k$, the width of the domain wall $1/m$, and
the size of the extra dimension $L$.  As will be shown in the next
subsection, we will be most interested in the case where $m \ll k$,
i.e. when the domain wall is thick in units of the curvature scale.
In this case, we find that $M_{P}^{2}$ is well approximated by
\bea
M_{P}^2 &\approx& \frac{2M^3_{5}}{m}  \times \sqrt{\frac{\pi m}{4\sqrt{2}k}} \, \tanh \left( \sqrt{\frac{2k}{m}} \times mL \right)~.
\label{mplapprox}
\eea
 From this expression we see that unless the extra dimension is so
small that $mL \ll \sqrt{m/k}$, the $\tanh$ is essentially equal
to one and we have $M^{2}_{P} \approx (\pi/\sqrt{2})^{1/2}\,
M^{3}_{5}/\sqrt{km} \approx 1.49 M^{3}_{5}/\sqrt{km}$.

For completeness, we also give an approximate expression that holds in
the narrow domain wall approximation, $m \gg k$:
\bea
M_{P}^2 &=& \frac{2M^3_{5}}{k} \int^{kL}_{0} \! dz \, \left[\cosh\left( \frac{m}{k} \sqrt{2} z \right) \right]^{-\sqrt{2} k/m}
\nonumber \\[0.5em]
&\approx& \frac{2M^3_{5}}{k} \left[ 2^{\sqrt{2} k/m - 1} \left( e^{-\sqrt{2} k/m} - e^{-2kL} \right) + \frac{k}{\sqrt{2}m} \right]~.
\label{mplapproxNarrow}
\eea
In the limit that $k/m \rightarrow 0$ we recover the RS result
$M_{P}^2 = ( 1 - e^{-2kL}) \, M^{3}_{5}/k$~\cite{Randall:1999ee}.

\subsubsection{Constraints from TeV Scale New Physics}
\label{ktilde1TeV}

We now explain how to further constrain the model parameters.
Imposing that the observed $M_{P}$ be reproduced, and for given
$k/M_{5}$, Eq.~(\ref{MPlanck}) determines $k$ as a function of $k/m$
and $mL$.  Fixing also $\tilde{k} \equiv k\,e^{-A(L)}$, one finds the
``required'' warp factor from $A(L) = \ln k/\tilde{k}$.  Comparing to
Eq.~(\ref{AL}) allows one to fix $mL$, for given $k/m$.~\footnote{For
$m \ll k$, Eq.~(\ref{mplapprox}) shows that $M_{P}$ becomes
essentially independent of $mL$ once $mL > \sqrt{m/k}$, while for
$m\gg k$ Eq.~(\ref{mplapproxNarrow}) shows that $M_{P}$ becomes
essentially independent of $mL$ for $mL > m/k$; in either of these
limits fitting $M_{P}$ effectively determines $k$ as a function of
$k/m$.  Then $mL$ is fixed as a function of $k/m$ from
Eq.~(\ref{AL}).} One can then find $\bar{\phi}_{0}/\phi_{0}$ from
Eq~(\ref{mL}).

\begin{figure}[t]
\centerline{ \hspace*{-0.5cm}
\includegraphics[width=0.48 \textwidth]{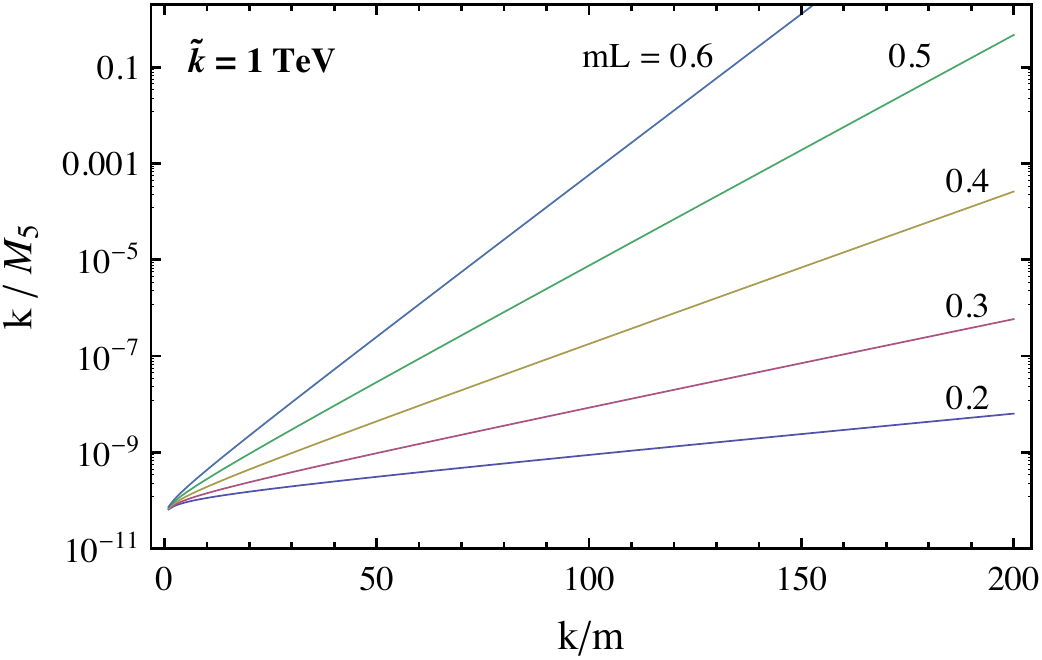}
\hspace*{0.3cm}
\includegraphics[width=0.48 \textwidth]{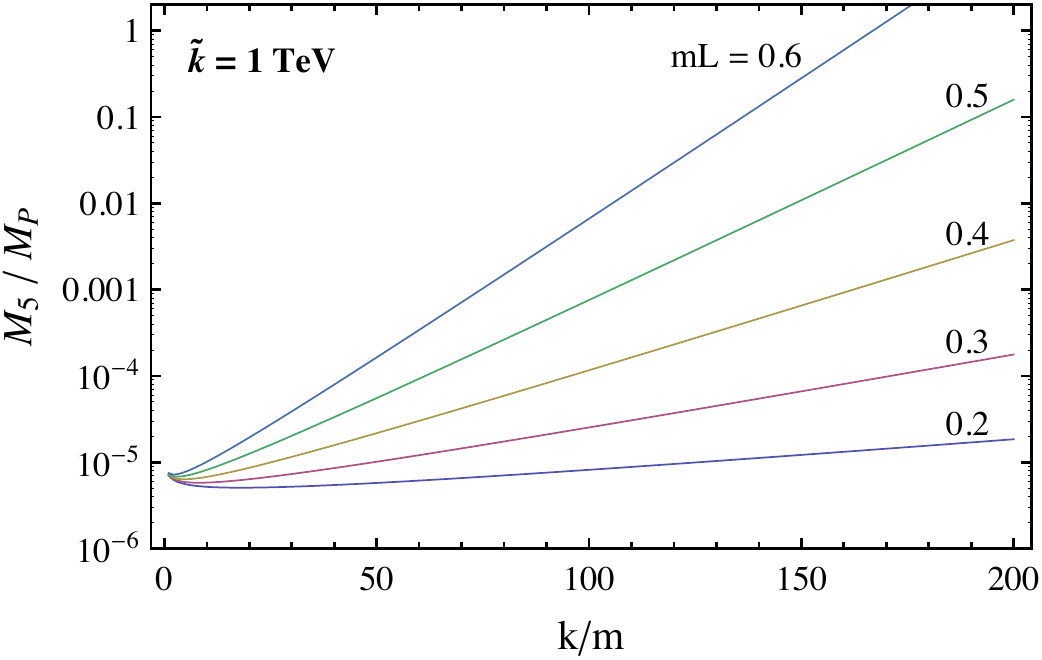}
}
\caption{\em $k/M_{5}$ (left panel) and $M_{5}/M_{P}$ (right panel) as
a function of $k/m$ for different values of $mL$, and for fixed
$\tilde{k} \equiv k\,e^{-A(L)} = 1~{\rm TeV}$.}
\label{fig:Scales_vs_kOvermAndmL}
\end{figure}
At this point the model is determined by $\epsilon \equiv k/M_{5}$ and
$k/m$.  In the left panel of Fig.~\ref{fig:Scales_vs_kOvermAndmL}, we
show lines of constant $mL$ in the $\epsilon{\rm -}(k/m)$ plane,
requiring that $M_{P} = 2.4 \times 10^{18}~{\rm GeV}$ and $\tilde{k} =
1~{\rm TeV}$.  This shows how to reach the regions of small ($k \ll
M_{5}$) and large ($k \lesssim M_{5}$) warping.  In the right panel,
we show what this implies for the ratio $M_{5}/M_{P}$ (recall that
$M_{5}$ is regarded as the fundamental scale in the theory, while
$M_{P}$ appears as an effective scale characterizing the interactions
of the zero-mode graviton).

\begin{figure}
\centerline{ \hspace*{-0.5cm}
\includegraphics[width=0.47 \textwidth]{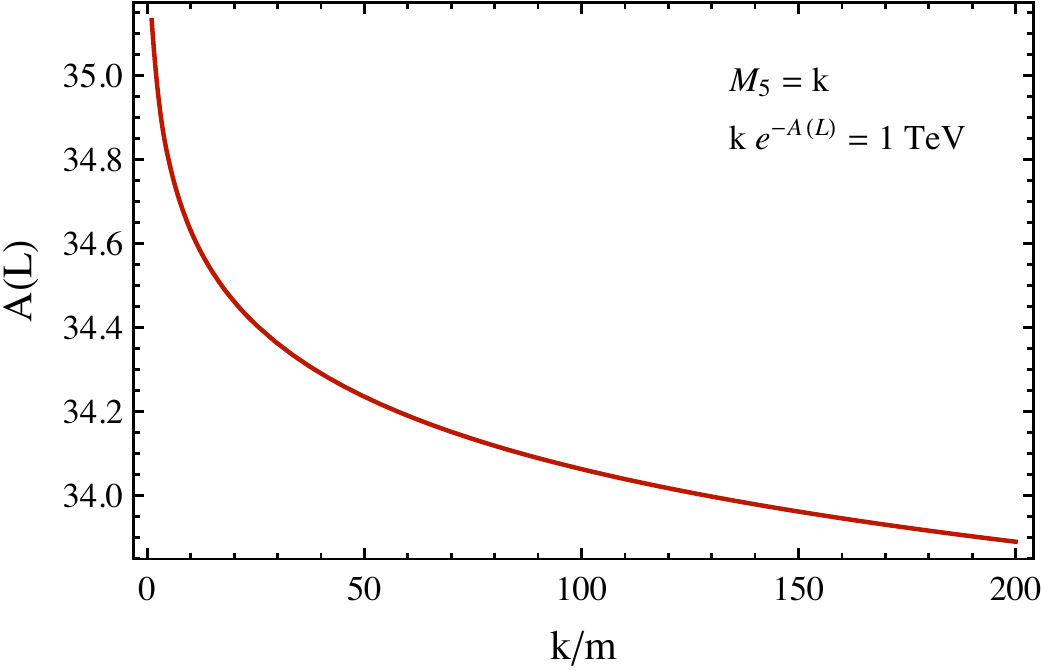}
\hspace*{0.5cm}
\includegraphics[width=0.455 \textwidth]{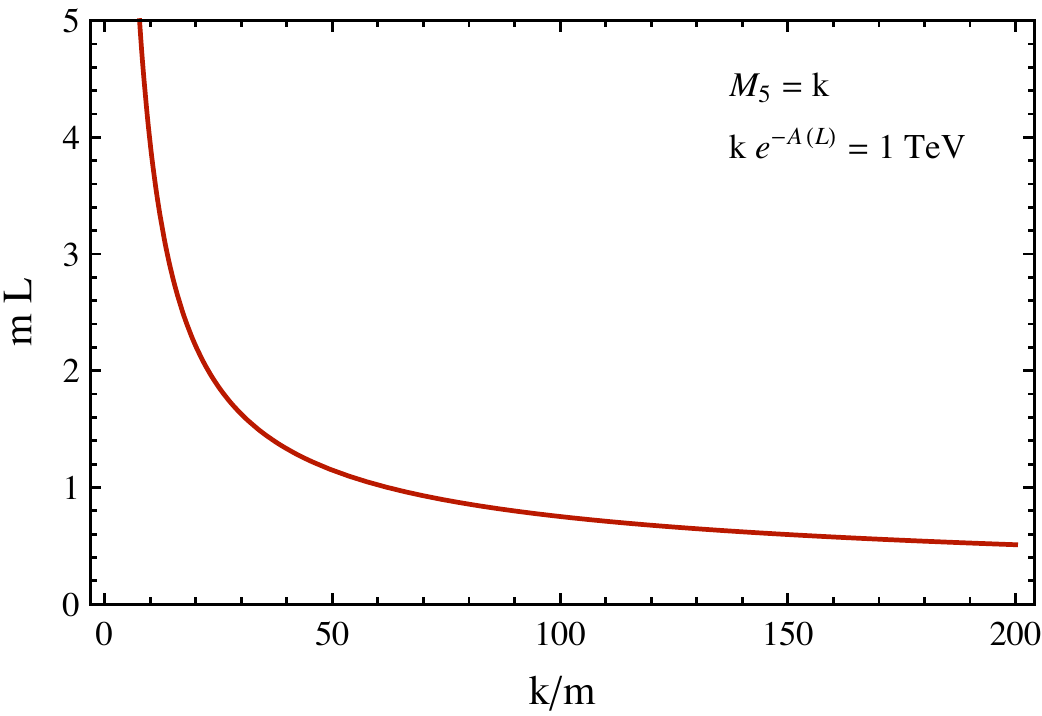}
}
\caption{\em $A(L)$ from Eq.~(\ref{AL}) (left panel), and $mL$ from
Eq.~(\ref{mL}) (right panel) as a function of $k/m$.  We impose
Eq.~(\ref{MPlanck}) assuming $M_{5} = k$ and $\tilde{k} \equiv
k\,e^{-A(L)} = 1~{\rm TeV}$.}
\label{fig:ALandmL}
\end{figure}

Of course, naturality considerations lead to the expectation that $k$
and $M_{5}$ should not be very different, and that $k \sim M_{5} \sim
M_{P}$ as in the Randall-Sundrum proposal.  In Fig.~\ref{fig:ALandmL}
we plot $A(L)$ and $mL$ as a function of $k/m$, assuming that $M_{5} =
k$.  We see that $A(L)$ remains essentially constant at $A(L) \sim
34{\rm -}35$.  The curvature scale varies between $(0.5-1.4) \times
10^{18}~{\rm GeV}$ in the shown range.  At the same time
$\bar{\phi}_{0}/\phi_{0}$ varies from $\bar{\phi}_{0}/\phi_{0} \sim
1.1$ for $k/m \sim {\rm few}$, down to $\bar{\phi}_{0}/\phi_{0} \sim
0.5$ for $k/m \sim 200$.  The right panel of Fig.~\ref{fig:ALandmL}
shows that $mL \gtrsim 1$ (we also point out that $35 \lsim kL \lsim
100$ in the range shown in the figures).  However, large values of
$mL$ require $\bar{\phi}_{0}/(\sqrt{2} \phi_{0}) \approx \pi/4$, so
that in general we expect $mL \sim {\cal O}(1)$, and therefore we need
$k/m \gg 1$ to generate the hierarchy.  This means that the physical
space is expected to be restricted to the ``central region'' of the
domain-wall solution.  As discussed in the previous subsection, in
this region the background solution is dominated by the quadratic term
in the scalar potential, and can be simply modeled by the linear
superpotential of Eq.~(\ref{LinearW}).

The small warping regime can be obtained for any value of $k/m$ with
$mL \lesssim 1$, as seen in the left panel of
Fig.~\ref{fig:Scales_vs_kOvermAndmL}.  For $k/m \sim 1$,
Eq.~(\ref{AL}) implies that $A(L) \sim {\cal O}(1)$, so that there is
virtually no warping.  In this case, the Planck-weak scale hierarchy
is directly set by $\tilde{k}/M_{P} \sim (k/M_{5}) \times
(M_{5}/M_{P}) \sim 10^{-15}$, as can be seen from
Fig.~\ref{fig:Scales_vs_kOvermAndmL}.~\footnote{Note that this gives a
scenario qualitatively similar to the UED models, where $1/L \sim k
\sim \tilde{k} \sim {\rm TeV}$, but $M_{5} \sim 10^{13}~{\rm GeV}$.
Since the couplings among KK modes are of the same order as those of
the 0-modes (recall $A(L) \sim 1$), the 4D EFT breaks down at ${\cal
O}(100) \times k$, as in UEDs.  However, as in warped scenarios, one
would expect that UV brane observables are under control up to scales
of order $M_{5}$.} For $k/m \gg 1$ one arrives at the interesting
situation where the Planck-weak scale hierarchy arises in part from a
sizable warping $1 < A(L) \ll {\cal O}(35)$ as well as from a
hierarchy associated with a small curvature $k \ll M_{5}$.

\subsubsection{The Limit of a Fat UV Brane: Benchmark Scenarios}
\label{Linear}

\begin{figure}
\centerline{ \hspace*{-0.5cm}
\includegraphics[width=0.47 \textwidth]{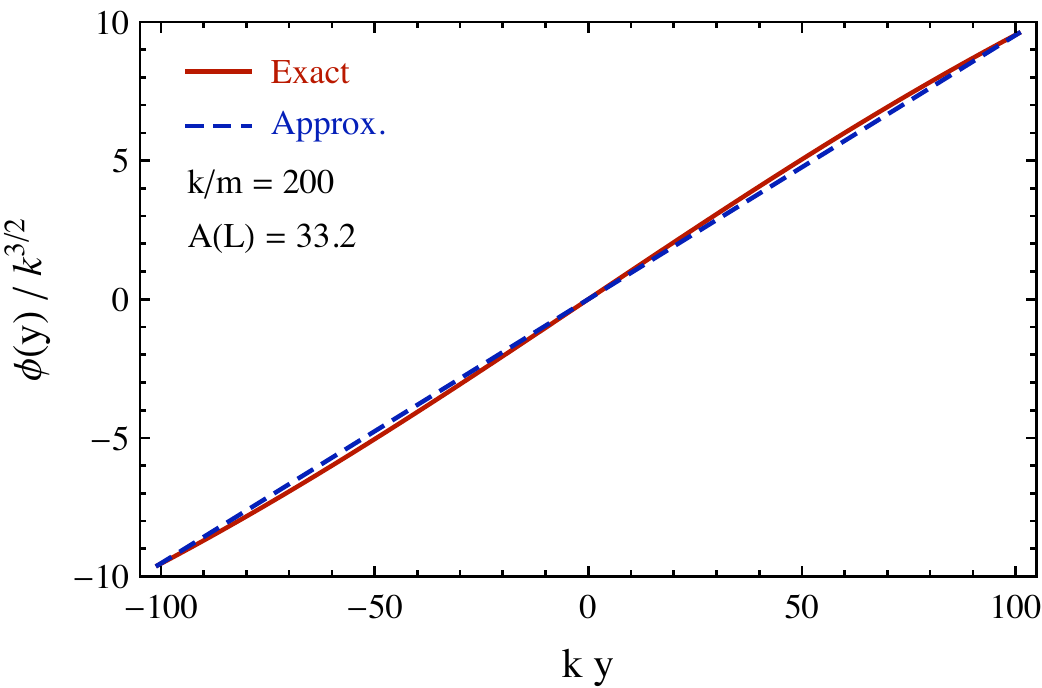}
\hspace*{0.5cm}
\includegraphics[width=0.47 \textwidth]{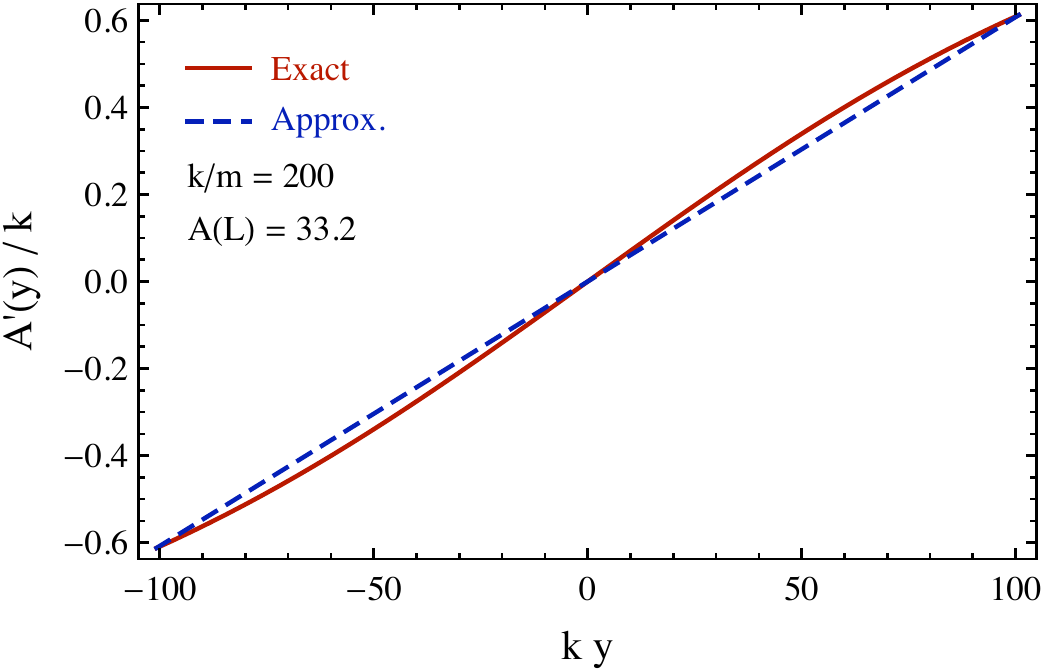}
}
\caption{\em Left panel: $\phi(y)$ in units of $k^{3/2}$ for the ``sine''
superpotential (solid, red line), and the approximation of
Eq.~(\ref{Approxphi}) (dashed, blue line).  Right panel: $A'(y)$ in
units of $k$ for the ``sine'' superpotential (solid, red line), and
the approximation of Eq.~(\ref{ApproxA}) (dashed, blue line).}
\label{fig:ApproximateProfiles}
\end{figure}

The previous considerations suggest that we focus on the regime $k/m
\gg 1$, corresponding to a ``fat UV brane''.  In this case, $\phi(y)$
and $A(y)$ can be approximated by linear and quadratic functions of
$y$ as follows:
\bea
\phi(y)/\phi_{0} &\approx& s_{\phi} y/L~,
\label{Approxphi}
\\[0.5em]
A(y) &\approx& s_{A} k y^2/(2L)~,
\label{ApproxA}
\eea
where $s_{\phi} = \sqrt{2} \tan^{-1} \left[\tanh
\left(\frac{mL}{\sqrt{2}} \right) \right]$ and $s_{A} = \tanh
\left(\sqrt{2} mL \right)$ in the model defined by the ``sine''
superpotential, Eq.~(\ref{sineW}).  In other models, $s_{\phi}$
and $s_{A}$ will have a different dependence on the microscopic
parameters, but will still be numbers of order one, so that
Eqs.~(\ref{Approxphi}) and (\ref{ApproxA}) provide a general
parameterization of ``fat brane scenarios'', which we will call
``the linear regime''.  The accuracy of the linear regime for $k/m
\gg 1$ can be seen in Fig.~\ref{fig:ApproximateProfiles}, where we
plot $\phi(y)$ and $A'(y)$ as a function of $ky$, for the exact
expressions (\ref{phivev}) and (\ref{ExplicitA}), as well as for
the approximations (\ref{Approxphi}) and (\ref{ApproxA}).  In the
following we will make extensive use of the approximate
expressions (\ref{Approxphi}) and (\ref{ApproxA}) to gain
intuition (and relatively accurate simple expressions) for the
spectrum and couplings of KK modes.

Here we will summarize the parameters for two ``benchmark scenarios''
that we will use in some of the numerical studies that follow.  Both
are based on the ``sine'' superpotential, i.e. on the profiles of
Eqs.~(\ref{phivev}) and (\ref{ExplicitA}).  As we will see in the
following, $k_{\rm eff} \equiv A'(L) (= s_{A} k)$ plays an important
role.
\begin{itemize}

\item
\textit{``Strong warping scenario''}: we take $M_{5} = k$, $\tilde{k}
= 2~{\rm TeV}$ and $k/m = 200$, apart from the phenomenological
constraints discussed above.  In this case, one finds $k \approx 5.2
\times 10^{17}~{\rm GeV}$, $A(L) \approx 33.2$, $mL \approx 0.50$,
$k_{\rm eff}L \approx 62$, $\bar{\phi}_{0}/\phi_{0} \approx 0.47$.
For reference, the radion decay constant, to be defined in the next
section, is $\Lambda_{r} = 4.4~{\rm TeV}$.

\item \textit{``Small warping scenario''}: we take $M_{5} = 2 \times
10^{8} k$, $\tilde{k} = 2~{\rm TeV}$ and $k/m = 200$, apart from the
phenomenological constraints discussed above.  In this case, one finds
$k \approx 1.8 \times 10^{5}~{\rm GeV}$, $A(L) \approx 4.5$, $mL
\approx 0.18$, $k_{\rm eff}L \approx 9$, $\bar{\phi}_{0}/\phi_{0}
\approx 0.18$.  The radion decay constant is $\Lambda_{r} = 2 \times
10^{16}~{\rm GeV}$.  This example lies approximately at the right end
of the blue, lower curve of Fig.~\ref{fig:Scales_vs_kOvermAndmL}.

\end{itemize}

Here we have taken a relatively large value for $k/m = 200$.  This is
only to make sure that we are in the linear approximation of
Eqs.~(\ref{Approxphi}) and (\ref{ApproxA}), since some of the simple
analytic expressions to be derived in the following sections rely on
being in the linear regime.  However, it is possible to take a smaller
hierarchy between $k$ and $m$, and we provide formulas for the general
case as well. We also point out that the most notable difference
between the strong and small warping scenarios enters through the
radion decay constant, $\Lambda_{r}$, and is therefore relevant for
radion or KK-radion phenomenology.  The specific ``small'' warping
scenario is inspired by a scenario for KK-radion dark matter, to be
presented in~\cite{Medina:2011qc}.

We should also add that there are further phenomenological constraints
on the KK scale from EW precision tests, and possibly from flavor and
CP violating observables.  However, such bounds depend on the
particular model under consideration.  Given that our emphasis in this
paper is on the general framework, we do not engage in the detailed
study of such constraints.  Nevertheless, we have chosen our ``strong
warping scenario'' in such a way that the KK gluon mass, $M_{\rm KK}
\approx 2.39 \, \tilde{k}_{\rm eff} \sim 3~{\rm TeV}$ (see caption to
Table~\ref{radionCoupStrongWarp} at the end of this paper), is roughly
consistent with the bounds studied in other warped scenarios.  For the
``small warping scenario'' we allow a smaller $M_{\rm KK} \approx 2.19
\, \tilde{k}_{\rm eff} \sim 1~{\rm TeV}$ (see caption to
Table~\ref{radionCoupSmallWarp}) since we expect the bounds to be
somewhat looser in that case (somewhat similar to the UED case).  We
stress, however, that the data we present in this work are largely
independent of the precise value of $M_{\rm KK}$, within a given
gravitational background, and should be useful when more detailed
studies of the bounds on the KK scale in given models are performed.

\section{Properties of the Radion KK Tower}
\label{RadionOddScalar}

We now turn to the scalar excitations associated with the background
described in the previous section, paying special attention to the
lightest KK-parity even scalar (the ``radion'') and the lightest
KK-parity odd scalar (that will turn out to be the LKP).

\subsection{Normalization and Radion Decay Constant}
\label{RadionNormalization}

We start by discussing the normalization of the radion/scalar KK modes
in general.  Following Ref.~\cite{Csaki:2000zn} we parameterize the 4D
physical scalar fluctuations (after gauge fixing) in the 5D
metric/bulk scalar system by
\bea
ds^2 &=& e^{-2A(y) - 2F(x,y)} \eta_{\mu\nu} dx^\mu dx^\nu - \left[ 1 + 2F(x,y)\right]^{2}dy^2~,
\label{MetricFluctuations}
\\ [0.5em]
\Phi(x,y) &=& \phi(y) + \varphi(x,y)~,
\label{ScalarFluctuations}
\eea
where $A(y)$ and $\phi(y)$ are the background profiles described in
the previous section, and
\bea
\varphi(x,y) &=& \frac{3 M_5^3}{\phi'(y)} \, e^{2A(y)} \partial_{y} \! \left[e^{-2 A(y)} F(x,y) \right]~.
\label{RadionScalarConnection}
\eea
In terms of the KK decomposition
\bea
F(x,y) = \sum_{n=0}^\infty \tilde{r}_n(x) F_n(y)~,
\label{RadionKKDecomposition}
\eea
one can show~\cite{Csaki:2000zn} that $F_{n}(y)$ obeys the ``KK
radion-scalar'' equation of motion~\footnote{Sometimes it is useful to
write this equation as $\phi^{\prime 2} \, \partial_{y} \!  \left\{
\frac{e^{2A}}{\phi^{\prime 2}} \, \partial_{y} \!  \left[ e^{-2A}
F_{n} \right] \right\} + \left( e^{2A} m_{n}^{2} - 2 A'' \right) F_{n}
= 0$, which suggests the relation between the existence of a massless
radion mode, with $F_{0} \propto e^{2A}$, and $A'' = 0$.}
\bea
F_{n}''-2A'F_{n}'-4A''F_{n} - 2\frac{\phi''}{\phi'} F_{n}' + 4 A' \frac{\phi''}{\phi'} F_{n} + e^{2A}
m_n^2 F_{n} &=& 0~,
\label{RadionEOM}
\eea
with boundary condition (in the limit that the $\Phi$ VEV is frozen at
the boundaries)
\bea
\left. (F_{n}' -2A' F_{n}) \right|_{\pm L} =0~,
\label{BCRadion}
\eea
guaranteeing the self-adjointness of Eq.~(\ref{RadionEOM}).  In order
to see the orthogonality relations for the KK wavefunctions that
follow from this, it is useful to remove the terms proportional to
$F_{n}'$ in Eq.~(\ref{RadionEOM}) by defining $\hat{F}_{n}(y) =
e^{-A(y)} F_{n}(y)/\phi'(y)$ so that:
\bea
\hat{F}_{n}'' - \left[ A^{\prime 2} + 3 A'' - 2 A' \, \frac{\phi''}{\phi'} + 2 \, \frac{\phi^{\prime\prime 2}}{\phi^{\prime 2}} - \frac{\phi'''}{\phi'} - e^{2A} m_n^2 \right]
\hat{F}_{n} &=& 0~.
\label{RadionEOMForOrthogonality}
\eea
The standard argument then leads to the orthogonality condition,
$\int_{-L}^{L} \!  dy \, e^{2A} \hat{F}_{m} \hat{F}_{n} \propto
\delta_{mn}$.  If we use the relation $\phi^{\prime 2} = 3 M_{5}^{3}
A''$, implied by the Einstein equations for the background, we get the
orthonormality relation for $F_{n}(y)$:
\bea
\int_{-L}^L \! dy \, \frac{F_m(y) F_n(y)}{A''(y)}
&=&  \frac{e^{-2A(L)}}{k_{\rm eff} m^{2}_{n}} \delta_{mn}~,
\label{RadionOrthonormality}
\eea
where the normalization of the wavefunctions, including the presence
of the KK radion masses $m_{n}$, was chosen for later convenience.

Replacing
Eqs.~(\ref{MetricFluctuations})-(\ref{RadionScalarConnection}) and
the KK decomposition of Eq.~(\ref{RadionKKDecomposition}) in the
action of Eq.~(\ref{GravityScalarAction}), one finds that the
Einstein-Hilbert term results in a contribution to the kinetic
terms of the radion KK-modes, $\tilde{r}_{n}$, given by $3
M^{2}_{5} \int \!  d^5 x \, e^{-2A} F_{m} F_{n} \, \partial_{\mu}
\tilde{r}_{m} \partial^{\mu} \tilde{r}_{n}$, where the derivatives
are contracted with the Minkowski metric.  Similarly, the scalar
part of the action gives a contribution to the KK radion kinetic
terms of $\int \!  d^5 x \, \frac{1}{2} \, e^{-2A} f_{n} f_{m} \,
\partial_{\mu} \tilde{r}_{m}
\partial^{\mu} \tilde{r}_{n}$, with $f_{n} = \frac{3 M_5^3}{\phi'} \,
e^{2A} \, \partial_{y} \!  \left(e^{-2 A} F_{n} \right)$.  After
integration by parts with respect to $y$, and using the KK radion
equation (\ref{RadionEOM}) together with the relation $\phi^{\prime 2}
= 3 M_{5}^{3} A''$, this generates a term that precisely cancels the
Einstein-Hilbert contribution, leaving behind only the terms
proportional to $m_{n}^2$.  The action Eq.~(\ref{GravityScalarAction})
at quadratic order in $\tilde{r}_n(x)$ then reads:
\bea
S &=& \frac{3}{2} M^{3}_{5} \, \sum_{m,n} \int \! d^4x \, m^{2}_{n}  \int_{-L}^{L} \! dy \, \frac{F_m(y) F_n(y)}{A''(y)} \, \partial_{\mu} \tilde{r}_{m} \partial^{\mu} \tilde{r}_{n} + \cdots~,
\nonumber \\[0.5em]
&=&
\Lambda^{2}_{r} \, \int \! d^4x \, \sum_n \frac{1}{2} \, \partial_\mu \tilde{r}_n(x) \partial^\mu \tilde{r}_n(x)  + \cdots~,
\label{RadionKineticTerms}
\eea
where we used the orthonormality relation
(\ref{RadionOrthonormality}), while the dots stand for higher order or
non-derivative terms, and contain the potential for $\tilde{r}_{n}$.
We also defined the radion decay constant
\bea
\Lambda_r = \sqrt{\frac{3 M_5^3}{k_{\rm eff}}} \, e^{-A(L)}~.
\label{rdecayconst}
\eea
In the case that Eq.~(\ref{mplapprox}) holds, i.e. for $m \ll k$, we
have $\Lambda_r \approx (2/s_{A})^{1/2} \left(\frac{m}{k}\right)^{1/4} M_P \,
e^{-A(L)}$.
\begin{figure}
\centerline{
\resizebox{9.cm}{!}{
\includegraphics{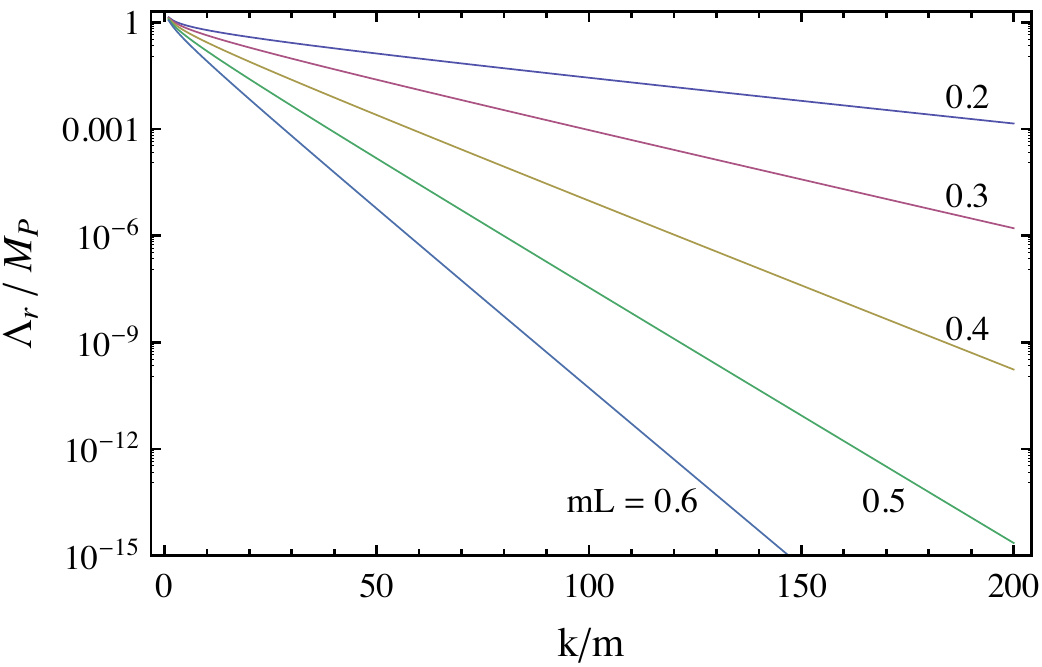}}
}
\caption{\em $\Lambda_{r}/M_{P}$ as a function of $k/m$ for several
values of $mL$.  Recall that $M_{P} = 2.4 \times 10^{18}~{\rm GeV}$ is
the reduced Planck mass.}
\label{fig:Lambdar}
\end{figure}
As remarked in Section~\ref{Scales}, the hierarchy between $k$ and
$M_{5}$ affects mostly the value of $\Lambda_{r}$, for fixed
$\tilde{k} = k\,e^{-A(L)}$.  As $\epsilon = k/M_{5}$ decreases,
$\Lambda_{r}$ increases, while for $k \sim M_{5}$, one has
$\Lambda_{r} \sim \tilde{k}$.  Thus, we see that the couplings of the
radion are very sensitive to the size of the 5D curvature.  In
Fig.~\ref{fig:Lambdar}, we show $\Lambda_{r}/M_{P}$ as a function of
$k/m$ for several values of $mL$, in the ``sine'' model of
Eqs.~(\ref{mL}) and (\ref{AL}).  This plot should be compared to
Fig.~\ref{fig:Scales_vs_kOvermAndmL}, but it is important to note that
$\Lambda_{r}/M_{P}$ is actually independent of $\tilde{k}$.

\subsection{Radion Mode}
\label{RadionMode}

In this subsection, we study the lightest excitations of
Eq.~(\ref{RadionEOM}).  Since in the limit that $A''(y) = 0$ there is
a zero-mode solution with wavefunction $F_{0} \propto e^{2A}$, it is
convenient to write in the general case
\bea
F_{0}(y) &\propto& e^{2A(y)+g(y)}~,
\label{RadionWavefunction}
\eea
so that the radion equation of motion and the boundary conditions
reduce to
\bea
g'' + {g'}^2+ 2 \left( A' - \frac{\phi''}{\phi'} \right) g' &=&  2 A'' -e^{2A}
m^{2}_{0}~,
\label{Eqg} \\
\left. g' \right|_{\pm L} &=& 0~.
\label{BCg}
\eea
Note that the differential equation is of first order in $g'$.  We can
obtain a good approximation to the ultralight mode with mass $m_{0}$
by neglecting the ${g'}^2$ term, in which case Eq.~(\ref{Eqg}) is
solved by
\bea
g'(y) &=& e^{-2A(y)} \phi'(y)^{2} \left[ 2 \int_{0}^{y} \! dz \, \frac{A''}{{\phi'}^{2}} \, e^{2A} - m^{2}_{0}   \int_{0}^{y} \! dz \, \frac{e^{4A}}{{\phi'}^{2}} \right]~,
\label{g'wavefunction}
\eea
where we chose the constant of integration so that $g'(0) = 0$ (since
the radion is even about $y = 0$).  The boundary condition (\ref{BCg})
then implies
\bea
m_{0} &=& \sqrt{\frac{2\int_{0}^{L} \! dz \, \frac{A''(z)}{k^{2}_{\rm eff}{\phi'(z)}^{2}} \, e^{2A(z)}}{\int_{0}^{L} \, \frac{dz}{{\phi'(z)}^{2}} e^{4A(z) - 2A(L)}}}~k_{\rm eff}\, e^{-A(L)}~,
\label{mRadionExact}
\eea
which is a rather general expression for the ``radion'' mass in terms
of $\phi(y)$ and $A(y)$.  Here we have expressed the result in terms
of $k_{\rm eff} = A'(L) = s_{A} k$.

For instance, for the ``strong'' (``small'') warping benchmark
scenarios defined at the end of Subsection~\ref{Linear}, we obtain
from Eq.~(\ref{mRadionExact}) that $m_{0} = 0.22 \,\tilde{k}_{\rm
eff}$ ($m_{0} = 0.67 \,\tilde{k}_{\rm eff}$), which agrees very well
with the numerical solution obtained without any approximations (here
$\tilde{k}_{\rm eff} \equiv k_{\rm eff}\, e^{-A(L)}$).  The neglected
${g'}^2$ term contributes negligibly to the mass because it is
comparable to the other terms in Eq.~(\ref{Eqg}) only for $y$'s near
the UV brane, but the wavefunction profile,
Eq.~(\ref{RadionWavefunction}), is exponentially localized near the
boundaries at $y = \pm L$.  Hence distortions near the UV region can
only give an exponentially small effect.

As explained in Section~\ref{Basic}, in order to solve the Planck/weak
scale hierarchy problem the size of the fifth dimension is often
stabilized in a region where $\phi(y)$ and $A(y)$ are well
approximated by linear and quadratic functions in $y$, as given in
Eqs.~(\ref{Approxphi}) and (\ref{ApproxA}), respectively.  Note that
the radion EOM, Eq.~(\ref{RadionEOM}), is approximately independent of
the scalar profile, since $\phi^{\prime\prime} \approx 0$ in this
limit.  In this case, we can write a simpler expression for $m_{0}$
that is independent of the scalar profile:
\bea
m_{0} &\approx& \sqrt{\frac{2}{k_{\rm eff} L} \frac{\int_{0}^{L} \! dz \, e^{2A(z)}}{\int_{0}^{L} \! dz \, e^{4A(z) - 2A(L)}}}~k_{\rm eff}\, e^{-A(L)}
\label{mRadionApprox1}
\\[0.5em]
&\approx& \frac{2}{\sqrt{k_{\rm eff}L}} \, k_{\rm eff}\, e^{-A(L)}~,
\label{mRadionApprox2}
\eea
where in the second line we used that when $A(y)$ is given by
Eq.~(\ref{ApproxA}), the ratio of integrals in
Eq.~(\ref{mRadionApprox1}) is very close to 2.~\footnote{If the linear
approximation is not excellent, one can obtain a pretty good
approximation to the radion mass by using $m_{0} \approx
\sqrt{\frac{2}{k_{\rm eff}L}} \, k_{\rm eff}\, e^{-A(L)}$, which only
assumes that, due to the strong localization of the inverse warp
factor near $y = L$ over a distance of order $1/k_{\rm eff}$, the
ratio of integrals in Eq.~(\ref{mRadionApprox1}) is of order one.} For
the ``strong'' (``small'') warping benchmark scenarios of
Subsection~\ref{Linear}, Eq.~(\ref{mRadionApprox1}) gives $m_{0} =
0.26 \,\tilde{k}_{\rm eff}$ ($m_{0} = 0.68 \,\tilde{k}_{\rm eff}$),
while Eq.~(\ref{mRadionApprox2}) gives $m_{0} \approx 0.25
\,\tilde{k}_{\rm eff}$ ($m_{0} \approx 0.67 \,\tilde{k}_{\rm eff}$).

For future reference, we also give an approximate expression for the
radion wavefunction, normalized according to
Eq.~(\ref{RadionOrthonormality}), in the limit that $A(y)$ is well
approximated by Eq.~(\ref{ApproxA}).  The fact that the radion mass is
parametrically smaller than the KK scale, $\tilde{k}_{\rm eff}$,
corresponds to the fact that $g(y)$ in Eq.~(\ref{RadionWavefunction})
is a small perturbation.  Therefore, setting $g(y) \approx 0$, and
using Eq.~(\ref{mRadionApprox2}) for $m_{0}$, we obtain
\bea
F_0(y) \approx e^{2[A(y)-A(L)]}~,
\label{F0Norm}
\eea
where we used $\int^z_{0} \!  dp \, e^{p^2} \approx
\frac{1}{2z} \, e^{z^2}$, which holds whenever $z \gg 1$.

\subsection{Lightest Odd-Scalar}
\label{RadionKKMode}

As already mentioned, the radion profile,
Eqs.~(\ref{RadionWavefunction}) or (\ref{F0Norm}), is exponentially
localized near the boundaries at $y = \pm L$.  Thus, after
normalization, its value at $y = 0$ is exponentially small (while
--being even-- its derivative vanishes at the origin).  One can then
see that by adjusting $m_{0}$ by an exponentially small amount one can
obtain a solution that vanishes at the origin, and therefore that
there is an odd mode with a mass exponentially close to the radion
mass given in Eq.~(\ref{mRadionExact}).  We can make the argument more
concrete by writing $F_{\rm odd}(y) = F(y) + \epsilon(y)$, where
$F(y)$ is the radion wavefunction given in
Eq.~(\ref{RadionWavefunction}).  Writing also $m^{2}_{n = 1} =
m^{2}_{0} + \delta m^{2}$, where $m_{0}$ is the radion mass given
above, Eq.~(\ref{RadionEOM}) becomes
\bea
\epsilon'' - 2 \epsilon' \left(A' + \frac{\phi''}{\phi'} \right) + \epsilon \left[ e^{2A} \left(m_{0}^{2} + \delta m^2 \right)  - 4A'' + 4 A' \frac{\phi''}{\phi'} \right] + e^{2A} \delta m^2 F = 0~.
\label{epsilonEOM}
\eea
We can approximately solve this equation in the limit where $\phi(y)$
and $A(y)$ are given by Eqs.~(\ref{Approxphi}) and (\ref{ApproxA}),
respectively, which corresponds to the limit we are interested in.  In
this case, the terms proportional to $\phi''$ vanish.  In addition,
using the approximate expression for $m_{0}$ given in
Eq.~(\ref{mRadionApprox2}) we see that $m_{0}^2 \, e^{2A}/(4A'') \sim
e^{-2[A(L) - A(y)]}$, which is much smaller than one except for $y
\approx L$.  Since, as we will see, also $\delta m^2 \ll m^2_{0}$, we
can focus on solving
\bea
\epsilon'' - 2 A' \epsilon'  - 4A'' \epsilon + e^{2A} \delta m^2 F &\approx& 0~,
\label{epsilonSimpleEOM}
\eea
with $A$ given by Eq.~(\ref{ApproxA}).  Its solution takes the
somewhat cumbersome form
\bea
\epsilon(y) &\approx& e^{2A(y)} \delta m^{2} \left\{ y \int_{y}^{L} \! dz_{2} \, F(z_{2}) \left[ 1 - 2 e^{2A(z_{2})} A'(z_{2}) \int_{z_{2}}^{L} \! dz_{1} \, e^{-2A(z_{1})}  \right] \right.
\nonumber \\[0.5em]
& & \hspace{2cm} \mbox{} -
\left. \left[ e^{-2A(y)}
- 2 A'(y) \int_{y}^{L} \! dz \, e^{-2A(z)} \right] \int_{y}^{L} \! dz \, z \, e^{2A(z)} F(z)
\right\}~,
\label{epsilonSoln}
\eea
where we imposed $\epsilon(L) = \epsilon'(L) = 0$.  The first
condition can be obtained by requiring that $F_{\rm odd}(L) = F(L)$ by
a simple overall rescaling of the odd wavefunction, and the second
follows from Eq.~(\ref{BCRadion}), together with the fact that the
radion wavefunction, $F(y)$, already satisfies this boundary
condition.  We emphasize that Eq.~(\ref{epsilonSoln}) assumes that
$A(y)$ is given by Eq.~(\ref{ApproxA}).  Requiring that $F_{\rm
odd}(0) = F(0) + \epsilon(0) = 0$ fixes $\delta m^{2}$ as
\bea
\delta m^{2} &\approx& \left\{\int_{0}^{L} \! dz \, z \, k^{2}_{\rm eff} \,e^{2[A(z) - A(L)]} \frac{F(z)}{F(0)}
\right\}^{-1} \, k^{2}_{\rm eff} \, e^{-2A(L)}
\label{OddMassApprox1}
\\ [0.5em]
&\approx& \frac{1}{k_{\rm eff}L} \, \frac{F(0)}{F(L)} \, k^{2}_{\rm eff} \, e^{-2A(L)}~,
\label{OddMassApprox2}
\eea
where we used $A(0) = A'(0) = 0$ to write Eq.~(\ref{OddMassApprox1}),
while Eq.~(\ref{OddMassApprox2}) holds due to the strong localization
of the radion wavefunction and the inverse warp factor near $y = L$
over a distance of order $1/k_{\rm eff}$.  This expression shows that
$\delta m^{2}/m_{0}^2$ is (exponentially) small due to the IR
localization of the radion wavefunction, $F(y)$.  Hence, the radion
and the lightest odd-scalar are exponentially degenerate in this
scenario.

\section{Fermions}
\label{sec:Fermion}

We analyze now the fermionic sector.  The Hermitian fermion action is
\bea
S_{\rm fermion} &=&
\int\! d^5x \sqrt{g} \left\{ \frac{i}{2}
\overline{\Psi} e^M_{A} \Gamma^{A} D_{M} \Psi - \frac{i}{2}
(D_{M}\Psi)^\dagger \Gamma^0 e^M_{A} \Gamma^{A} \Psi - y_{\Psi} \Phi
\overline{\Psi} \Psi \right\}~,
\label{FermionAction}
\eea
where $\Gamma^{A} = (\gamma^\mu, -i \gamma_{5})$ are the flat space
Dirac gamma matrices in 5D space, $e^M_{A}$ is the f\"unfbein, $D_{M}$
is the covariant derivative with respect to the gauge symmetry as well
as general coordinate and local Lorentz transformations~\footnote{For
a diagonal metric of the form $ds^2 = a(y)^2 \eta_{\mu\nu} dx^\mu
dx^\nu - b(y)^2 dy^2$ the spin connection in $D_{M}$ cancels out in
the fermion action, Eq.~(\ref{FermionAction}).}, and $y_{\Psi}$ is a
Yukawa coupling (with mass dimension $-1/2$).  In the scalar
background $\langle\phi\rangle$ given in Eq.~(\ref{phivev}) the Yukawa
interaction gives rise to a $y$-dependent mass term for the fermion,
$m_{D} \equiv y_{\Psi} \langle \Phi \rangle$. 

Note that the Yukawa interaction above is consistent with the $Z_{2}$
transformation $\Phi(y) \to -\Phi(-y)$ provided either $\Psi_{L}(y)
\to \Psi_{L}(-y)$ and $\Psi_{R}(y) \to -\Psi_{R}(-y)$ or viceversa.
This matches with the fact that, at a given KK level, the $y \to -y$
transformation properties of the wavefunctions $f_{L}^{n}(y)$ and
$f_{R}^{n}(y)$ are opposite, but the associated 4D fields are
nevertheless assigned the same KK parity (even for even $n$ and odd
for odd $n$, independent of chirality).

\subsection{Equations of Motion and Kaluza-Klein Decomposition}
\label{sec:KKFermion}

Considering the action to quadratic order in the fields, the KK
decomposition for each 4D chirality of the 5D fermion
reads~\footnote{We use the convention that $\gamma_{5} \Psi_{L,R} =
\mp \Psi_{L,R}$.}
\begin{equation}
\Psi_{L,R}(x,y)= \frac{e^{\frac{3}{2}A(y)}}{\sqrt{2L}}
\sum_{n=0}^{\infty}\psi^{n}_{L,R}(x)
f^n_{L,R}(y)~,
\label{5Dfermion}
\end{equation}
where the fermion profiles obey the orthonormality conditions
\bea \frac{1}{2L} \int_{-L}^L \! dy \, f_{L,R}^n(y) f_{L,R}^m(y)
&=& \delta_{nm}~,
\label{fermonNormalization}
\eea
and the first order equations
\begin{eqnarray}
\left(\partial_{y}+m_{D} - \frac{1}{2} A' \right) f^n_{L}&=&m_{n}e^{A}f^n_{R}~,
\label{f_lFOeq}
\\[0.5em]
\left(\partial_{y}-m_{D} - \frac{1}{2} A' \right) f^n_{R}&=&-m_{n}e^{A}f^n_{L}~.
\label{f_rFOeq}
\end{eqnarray}
The boundary conditions at $y=-L$ and $y=L$ are given either by
\begin{eqnarray}
\left. {f^{n}_L}'+\left(m_{D}-\frac{1}{2} A' \right)f^{n}_{L} \right|_{y = \mp L} &=& 0~,
\quad \quad  f^{n}_{R}(\mp L) ~ = ~ 0~,
\label{BCsfL}
\end{eqnarray}
or
\begin{eqnarray}
\left. {f^{n}_R}'-\left(m_{D}+\frac{1}{2} A' \right)f^{n}_{R} \right|_{y = \mp L} &=&0~,
\quad \quad f^{n}_{L}(\mp L) ~ = ~ 0~,
\label{BCsfR}
\end{eqnarray}
where $'=\partial_{y}$.  These give rise to a chiral zero-mode sector:
Eq.~(\ref{BCsfL}) selects a left-handed (LH) zero-mode while
Eq.~(\ref{BCsfR}) allows only a right-handed (RH) one.  The zero-mode
solutions are explicitly given by
\bea
f_{L,R}^0(y) &=& N \exp \left(\frac{1}{2} A(y) \mp \int^y_{0} \! dz \, m_D (z) \right)~,
\label{zeromodes}
\eea
where $N$ is determined by Eq.~(\ref{fermonNormalization}).  These
solutions are even about $y = 0$ since $A(y)$ is even and the integral
of the odd function $m_{D}(z)$ is also even.

\begin{figure}
\centerline{ \hspace*{-0.5cm}
\includegraphics[width=0.47 \textwidth]{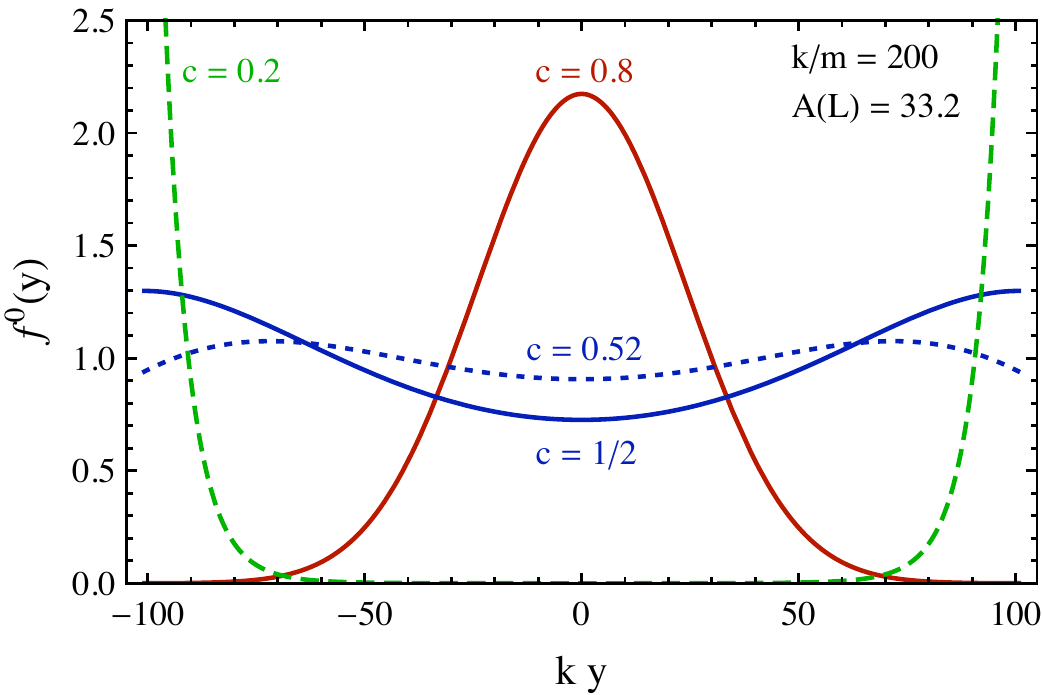}
\hspace*{0.5cm}
\includegraphics[width=0.47 \textwidth]{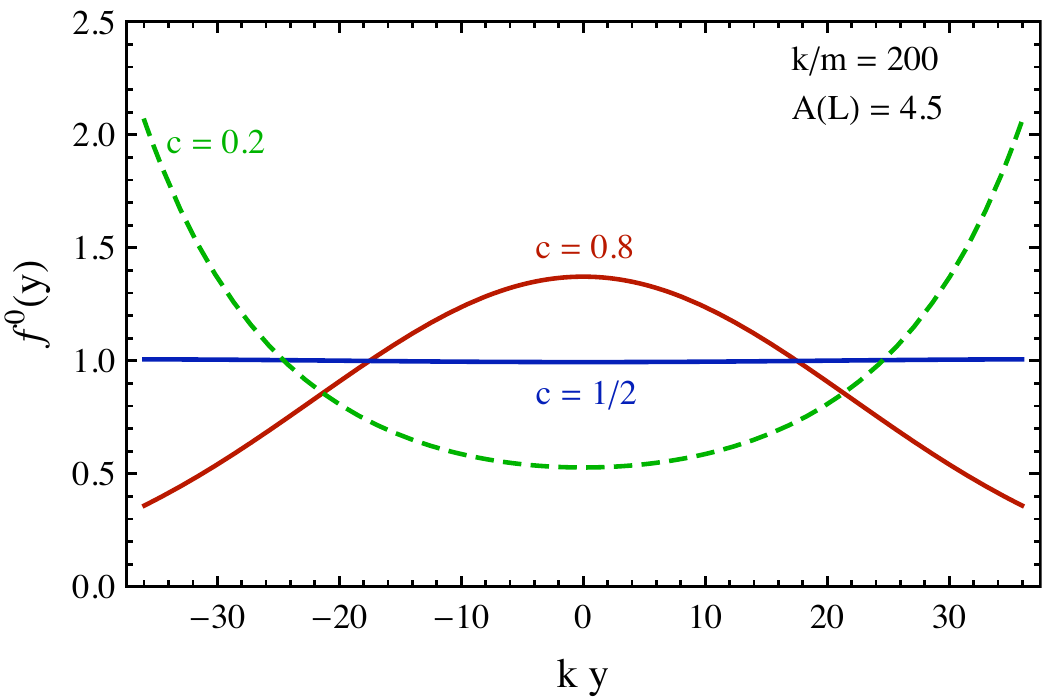}
}
\caption{\em Fermion zero-mode wavefunctions, Eq.~(\ref{zeromodes}),
for the ``strong warping'' (left panel) and ``small warping'' (right
panel) scenarios.  The $c$-parameter is defined by
Eq.~(\ref{cdefinition}).  Note that the Randall-Sundrum intuition
approximately applies, but there is no symmetry between UV and IR
localized wavefunctions due to the approximately Gaussian warp factor.
The dynamical UV brane is at $y = 0$, while $y=\pm L$ correspond to IR
boundaries.}
\label{fig:fermion0modes}
\end{figure}

We note that when $m_{D}(y)$ and $A'(y)$ are proportional to each
other, which is the case when $\phi(y)$ and $A(y)$ are approximately
given by Eqs.~(\ref{Approxphi}) and (\ref{ApproxA}), the fermion
zero-mode wavefunctions further simplify.  In this case, we define a
dimensionless parameter $c$ by
\bea
m_{D}(y) &=& \pm c A'(y)~,
\label{Conformal}
\eea
where we take the upper (lower) sign when the 5D fermion has a LH (RH)
zero mode (i.e. when the boundary conditions (\ref{BCsfL}) or
(\ref{BCsfR}) hold, respectively).  In our case, we have the explicit
relation
\bea
c &=& \pm y_{\Psi} (s_{\phi}/s_{A})(\phi_{0}/k)~,
\label{cdefinition}
\eea
where $s_{\phi}$ and $s_{A}$ were defined in Eqs.~(\ref{Approxphi})
and (\ref{ApproxA}).  In the limit that the proportionality of
Eq.~(\ref{Conformal}) holds, the zero-mode profiles of
Eq.~(\ref{zeromodes}) take the simple form
\bea
f^0(y) &=& N_0 \, e^{-\left( c - \frac{1}{2} \right) A(y) }~,
\label{zeromodeSimple}
\eea
independently of the chirality of the zero-mode sector.  This shows
that for $c \approx 1/2$ the zero-mode profiles are (approximately)
flat, while for $c \gtrsim 1/2$ ($c \lesssim 1/2$) they are localized
near the UV brane (IR boundaries).  Even in cases where
Eq.~(\ref{Conformal}) does not hold exactly, it is useful to formally
exchange the Yukawa coupling $y_{\Psi}$ for a $c$-parameter, defined
by Eq.~(\ref{cdefinition}), and we shall do so from now on.  The
model-dependent constants, $s_{\phi}$ and $s_{A}$ can be easily
estimated in the ``central region'' of the kink scalar profile [and
were given after Eq.~(\ref{ApproxA}) for the ``sine'' superpotential,
Eq.~(\ref{sineW})].  In Fig.~\ref{fig:fermion0modes}, we show the
exact fermion zero-mode wavefunctions for several values of $c$ to
illustrate the situation in the benchmark scenarios defined in
Subsection~\ref{Linear}, with $k \gg m$.  This allows us to
characterize the physics of localization in close analogy to the
standard treatment of bulk fermions in the Randall-Sundrum scenario.
But note that, in general, Eq.~(\ref{zeromodes}) does not admit an
exactly flat (``conformal'') zero-mode solution, since the
$y$-dependence of the two terms in the exponent cannot cancel for all
$y$'s unless Eq.~(\ref{Conformal}) holds.\footnote{This also shows
that our background would break supersymmetry in a ``hard way'', since
the gaugino zero-mode profile, hence its couplings, cannot exactly
match those of the gauge field.} Also, the UV-localized wavefunctions
have a Gaussian-like profile (see
also~\cite{ArkaniHamed:1999dc,Kaplan:2001ga}), while the IR-localized
wavefunctions have an exponential-like profile \'a la RS.
Nevertheless, it is interesting that we end up dynamically in a limit
where the RS intuition qualitatively holds.  The approximate
expression for the zero-mode profiles, Eq.~(\ref{zeromodeSimple}),
will be very useful in the following sections.

\begin{figure}[t]
\centerline{ \resizebox{9cm}{!}{
\includegraphics{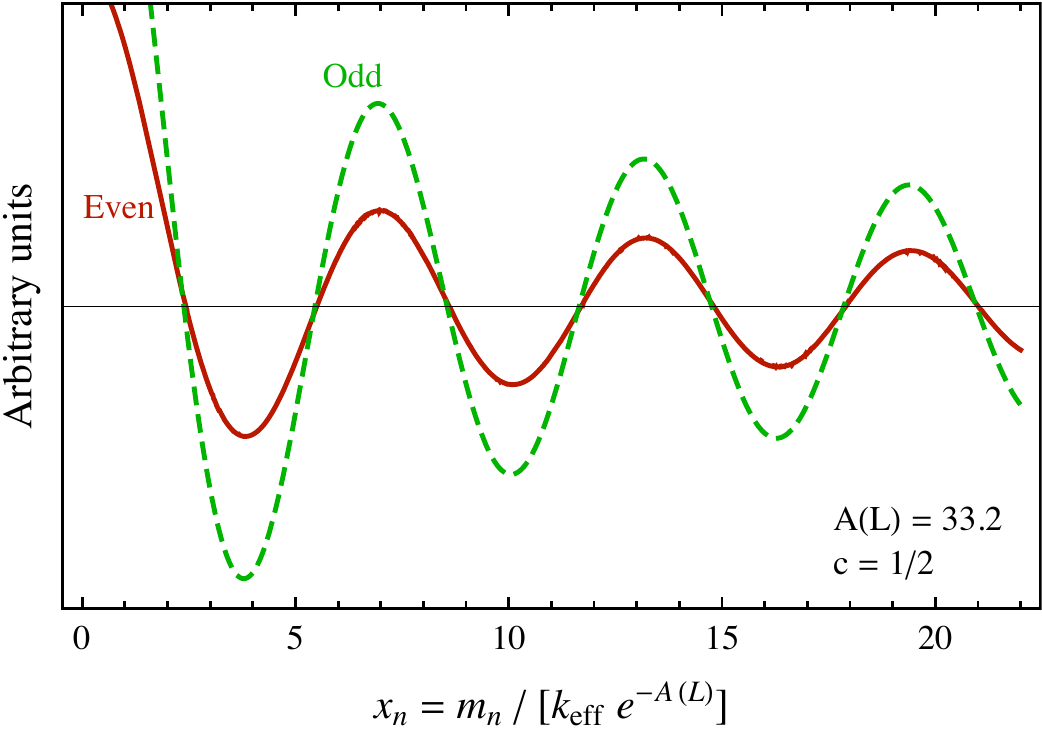}}
}
\caption{\em We plot $f_{R}(0)$ (solid, red curve) and $f_{L}(0)$
(dashed, green curve) as a function of $x_{n} = m_{n}/[k_{\rm
eff}\,e^{-A(L)}]$, for the boundary conditions of Eq.~(\ref{BCsfL})
[which select a LH zero-mode].  The zeros of $f_{R}(0)$ (which imply
zeros of $f_{L}'(0)$) correspond to the even mode LH solutions, while
those of $f_{L}(0)$ correspond to the odd mode LH solutions.  We
choose $c = 1/2$ [approximately flat fermion zero-mode in the linear
regime of Eqs.~(\ref{Approxphi}) and (\ref{ApproxA})], so that the
roots also approximately correspond to those of gauge fields (see
text).  In this example, the roots are at $x^{\rm even}_{n} \approx
\{0,2.43,5.51,8.60,11.70,14.80, \ldots\}$ and $x^{\rm odd}_{n} \approx
\{2.38,5.46,8.55,11.65,14.75, \ldots\}$.  Notice the near degeneracy
between the even and odd spectra.}
\label{fig:roots}
\end{figure}

Going back to the heavy KK-mode fermion solutions, combining
Eqs.~(\ref{f_lFOeq}) and (\ref{f_rFOeq}) we obtain the decoupled
system of second order differential equations:
\begin{eqnarray}
{f^{n}_{L}}'' - 2 A' {f^{n}_{L}}' + \left[ \frac{3}{4} A'^2 - \frac{1}{2} A'' -
m_{D} A' + m_{D}' -m_D^2 + e^{2A} m_n^2 \right] f^n_{L} &=& 0~,
\label{fleq}
\\[0.5em]
{f^{n}_{R}}''-2 A' {f^{n}_{R}}' + \left[ \frac{3}{4} A'^2 - \frac{1}{2} A'' +
m_{D} A' - m_{D}' -m_D^2 + e^{2A} m_n^2 \right] f^n_{R} &=& 0~.
\label{freq}
\end{eqnarray}
In the absence of closed solutions for our chosen $A(y)$ and
$m_{D}(y)$, one can find the spectrum (and wavefunctions) numerically
by the ``shooting method''.\footnote{From a numerical perspective, it
is easier to solve for $e^{-\frac{1}{2} A(y)} \, f^{n}_{L,R}$.} For
instance, concentrating on $f_{L}^n$, one can integrate numerically
Eq.~(\ref{fleq}) by starting at $y = -L$ with $f_{L}^n(-L) = 1$ and
its derivative given by the first equation in~(\ref{BCsfL}).  Varying
$x_{n} = m_{n}/[k_{\rm eff}\,e^{-A(L)}]$, one can find those values of
$x_{n}$ for which ${f_{L}^n}'(0) = 0$.  These correspond to the even
mode solutions.  Similarly, the values of $x_{n}$ that make
$f_{L}^n(0) = 0$ determine the odd mode solutions.  In practice, it is
easier to find the even LH spectrum by solving Eq.~(\ref{freq}) for
$f_{R}^n$ and requiring $f_{R}^{n}(0) = 0$, which by the first order
Eq.~(\ref{f_lFOeq}) implies ${f_{L}^n}'(0) = 0$.  This only misses the
LH zero-mode solution, that we know explicitly from
Eq.~(\ref{zeromodes}).  We show in Fig.~\ref{fig:roots} the result of
the above procedure for the ``strong warping'' benchmark scenario
defined in Subsection~\ref{Linear}, taking $c = 1/2$, so that the
fermion zero mode is (approximately) flat.  We use the exact profiles
for $A(y)$ and $\phi(y)$ to generate the figure.  However, to the
extent that Eqs.~(\ref{Approxphi}) and (\ref{ApproxA}) hold, this
example corresponds to the ``conformal'' case, and the solutions for
the $x_{n}$ approximately coincide with those of (unbroken) gauge
fields.\footnote{Assuming vanishing brane-localized kinetic
terms~\cite{BKTs}.} Indeed, one can check that the gauge spectrum
coincides with the one shown in the figure at the percent level.  We
note from the figure the high degree of degeneracy between the even
and odd KK modes.  This is due to the fact that the KK-mode
wavefunctions are highly peaked near the IR boundaries, hence small
changes in $x_{n}$ allows one to change from ${f_{L}^n}'(0) = 0$ to
$f_{L}^n(0) = 0$, turning an even mode into an odd mode (the same
effect discussed for the radion/odd-scalar in
Section~\ref{RadionOddScalar}).  The degree of degeneracy increases as
$A(L)$ increases.

As discussed above, there is always an even zero-mode.  In addition,
in certain regions of parameter space, there are light vector-like
odd-modes that can be treated analytically, as we discuss in the next
subsection.

\subsection{Light Vector-like Odd Modes}
\label{sec:LightFermion}

For concreteness, in the following we concentrate on the case of a
left-handed zero mode, i.e. we impose the boundary conditions of
Eq.~(\ref{BCsfL}) [our results for the light mass and its wavefunction
also hold when the zero-mode is right-handed, by a simple exchange of
the labels $L \leftrightarrow R$].  In the case that the zero-mode
fermion is exponentially localized near the IR boundaries, we expect
to find an ultralight odd-fermion solution (for the same reasons as
for the radion case discussed above).  To find this small mass we
write the wavefunction for the LH chirality of this odd-fermion (the
first KK-mode), on the interval $[0,L]$, as $f_L^1(y) = f_L^0(y) +
\epsilon(y)$, where $f_L^0$ is the (even) zero-mode profile given in
Eq.~(\ref{zeromodes}) [$f_L^1(y)$ is completely fixed on the interval
$[-L,0]$ from the fact that it is odd about $y=0$].  One then finds
that Eq.~(\ref{fleq}) becomes
\bea \epsilon'' - 2 A' \epsilon' + \left[ \frac{3}{4} A'^2 - \frac{1}{2} A'' -
m_{D} A' + m_{D}' -m_D^2 + e^{2A} m_1^2 \right] \epsilon  + e^{2A} m_1^2 f_L^0 &=& 0~,
\label{Eqepsilon1}
\eea
and requiring that the zero mode and the first excited mode have the
same value at $L$ we have the boundary conditions $\epsilon(L) =
\epsilon'(L) = 0$.

We give an approximate solution for $\epsilon(y)$ in the limit that
$\phi(y)$ and $A(y)$ are approximately given by Eqs.~(\ref{Approxphi})
and (\ref{ApproxA}), respectively, and therefore Eq.~(\ref{Conformal})
approximately holds (with the upper sign).  Neglecting the $e^{2A}
m_1^2 \epsilon$ term (which is smaller than the other terms
proportional to $\epsilon$ provided $m_{1} \ll k\, e^{-A(L)}$),
Eq.~(\ref{Eqepsilon1}) can be written as
\bea
e^{\left( c - \frac{1}{2} \right) A } \partial_{y} \left\{ \partial_{y}\left[e^{\left( c - \frac{1}{2} \right)A} \, \epsilon
\right]
e^{-\left(2c + 1 \right) A} \right\}
&=& - m_1^2 f_L^0~.
\label{Eqepsilon2}
\eea
This can be integrated immediately to give
\bea
\epsilon(y)
&=& -m_1^2 \, e^{- \left( c - \frac{1}{2}\right)A(y)}
\int_y^L \! dz_2 \, e^{\left(2c + 1 \right) A(z_2)}
\int_{z_2}^L \! dz_1 \,
e^{- \left( c - \frac{1}{2} \right) A(z_1) } f_L^0(z_1)~,
\label{epsilonFermion}
\eea
where we imposed the boundary conditions given after
Eq.~(\ref{Eqepsilon1}).  Requiring that $f_L^1(0) = f_L^0(0) +
\epsilon(0) = 0$ determines the lightest odd-fermion mass as
\bea m_1^2 &\approx& \left\{ \int_0^L \! dz_2 \, k_{\rm eff} \, e^{\left(2c + 1 \right) A(z_2) - A(L)}
\int_{z_2}^L \! dz_1 \, k_{\rm eff} \,
e^{- \left( c - \frac{1}{2} \right) A(z_1) - A(L) } \, \frac{f_L^0(z_1)}{f_L^0(0)}
\right\}^{-1}  k^2_{\rm eff} e^{-2A(L)}~,
\hspace{5mm}
\label{m1FermionApprox1}
\eea
where we used $A(0) = 0$.  This form shows that the lightness of this
odd mode is associated with $f_L^0(0)/f_L^0(L) \ll 1$.  Indeed, since
in the present limit the zero-mode profile is well approximated by
Eq.~(\ref{zeromodeSimple}), we can write
\bea
m_1^2 &\approx& \left\{ \int_0^L \! dz_2 \, k_{\rm eff} \, e^{\left( 2c + 1 \right) A(z_2) - A(L)}
\int_{z_2}^L \! dz_1 \, k_{\rm eff} \,
e^{- \left( 2c - 1 \right) A(z_1) - A(L) }
\right\}^{-1}  k^2_{\rm eff} e^{-2A(L)}~.
\label{m1FermionApprox2}
\eea
This expression allows us to understand the suppression in $m_{1}$
compared to the natural KK scale $k_{\rm eff}\,e^{-A(L)}$ when $c <
1/2$, i.e. when the zero-mode is IR localized and we expect a light
odd-fermion mode [thus validating the approximation made above of
neglecting the $e^{2A} m_1^2 \epsilon$ term in
Eq.~(\ref{Eqepsilon2})].  In such a case, and using the approximation
(\ref{ApproxA}) for $A(y)$, the factor $e^{- \left( 2c - 1 \right)
A(z_1)}$ is localized near the IR boundary over a distance of order
$[(1-2c) k_{\rm eff}]^{-1}$, and therefore the $z_{1}$ integration can
be restricted to the region $[L - \{(1 - 2c) k_{\rm eff} \}^{-1},L]$.
To proceed further, it is useful to distinguish two cases: \\

\vspace{-3mm}
\noindent
$\bullet$ $-1/2 < c < 1/2$: the factor $e^{\left( 2c + 1 \right)
A(z_2)}$ is also localized towards the IR boundary, over a distance of
order $[(2c + 1) k_{\rm eff}]^{-1}$.  We can therefore restrict the
integral over $z_{2}$ to the region $[L - \{(1 + 2c) k_{\rm eff}
\}^{-1},L]$.  With this restriction over $z_{2}$ we find that the
effective support of the $z_{1}$ integration is of order $(2k_{\rm
eff})^{-1}$, in the vicinity of $z_{1} = L$.  Evaluating all the
exponentials at $L$, we then get
\bea
m_1 &\approx&  \sqrt{2(1 + 2c)} \, k_{\rm eff} \, e^{-A(L)}~,
\label{m1FermionSmallc}
\eea
where we recall that $k_{\rm eff} = A'(L) = s_{A} k$.
\\

\vspace{-3mm}
\noindent
$\bullet$ $c < -1/2$: the factor $e^{\left( 2c + 1 \right) A(z_2)}$ is
now localized near $y = 0$.  The integral over $z_{1}$ can then be
evaluated from $0$ to $L$ for all $z_{2}$, giving a factor $[(1-2c)
k_{\rm eff}]^{-1}e^{-2c A(L)}$.  For $|c| \sim {\cal O}(1)$, the integral
over $z_{2}$ can then be estimated as an average and gives $L\,
e^{-A(L)} (1/2) [e^{(2c+1)A(L)} + e^{(2c+1)A(0)}]$.  Putting these
together, we have
\bea
m_1 &\approx&  \sqrt{\frac{2 (1 - 2c)}{k_{\rm eff} L} \frac{1}{1+e^{-(2c+1)A(L)}}} \, k_{\rm eff} \, e^{-A(L)}~,
\label{m1FermionNegc}
\eea
where we used $A(0) = 0$.  This shows that for extreme localization of
the zero-mode solution, i.e. for $2c + 1$ very negative, the lightest
odd-fermion mode has an exponentially small mass, where
$m_{1}/(k_{\rm eff}\,e^{-A(L)}) \sim e^{(c+\frac{1}{2}) A(L)}$.

\begin{figure}
\centerline{
\resizebox{9.cm}{!}{
\includegraphics{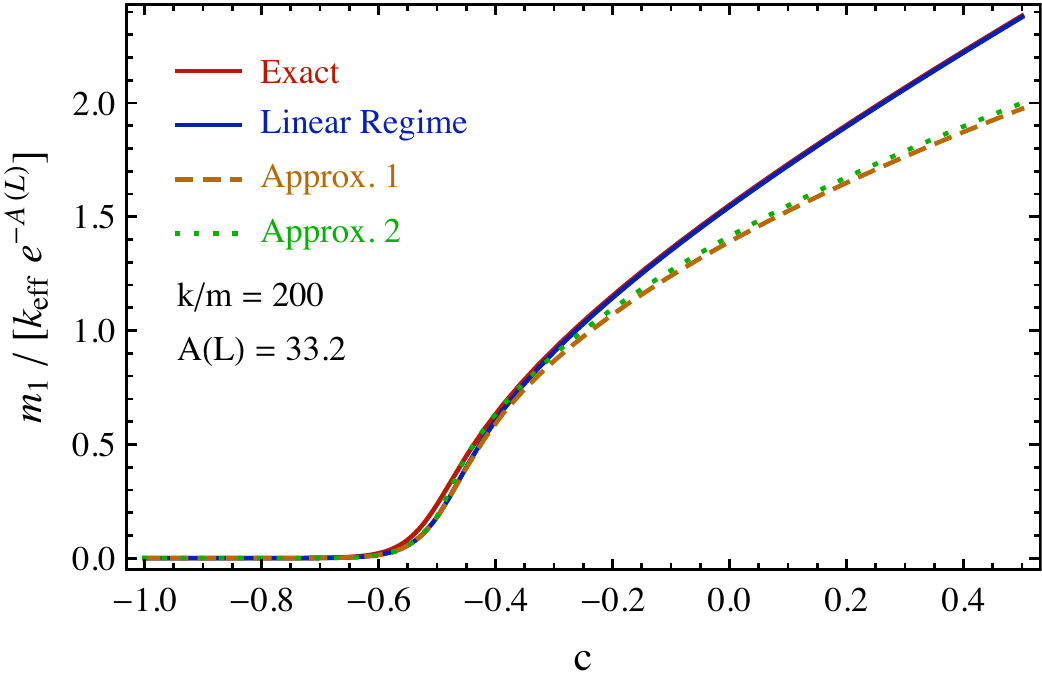}} }
\caption{\em Lightest odd-fermion mass (in units of $k_{\rm
eff}\,e^{-A(L)}$) as a function of c.  The solid (red) line
corresponds to the numerical result with the exact scalar and warp
factor profiles of Eqs.~(\ref{phivev}) and (\ref{ExplicitA}).  The
solid (blue) line corresponds to the numerical solution but with
$\phi(y)$ and $A(y)$ given by Eqs.~(\ref{Approxphi}) and
(\ref{ApproxA}) (the ``linear regime'').  The dashed (brown) line
corresponds to the approximation of Eq.~(\ref{m1FermionApprox2})
(``Approx.  1'').  The dotted (green) line corresponds to
Eqs.~(\ref{m1FermionSmallc}) and (\ref{m1FermionNegc}) (``Approx.
2'').}
\label{m1fermion:fig}
\end{figure}
In Fig.~\ref{m1fermion:fig}, we show the lightest odd-fermion mass (in
units of $k_{\rm eff}\,e^{-A(L)}$) as a function of c for the
approximation given in Eq.~(\ref{m1FermionApprox2}) with $A(y) =
k_{\rm eff} y^2/(2L)$ [dashed, brown line], and for
Eqs.~(\ref{m1FermionSmallc}) and (\ref{m1FermionNegc}) in the regions
$-1/2 < c < 1/2$ and $c < -1/2$, respectively [dotted, green
line].~\footnote{The approximation of Eq.~(\ref{m1FermionSmallc})
fails at $c = -1/2$, since the factor $e^{\left( 2c + 1 \right)
A(z_2)}$ becomes constant.  However, numerically the approximation
works rather well down to $c \simeq -0.48$.  For the dotted green line
in Fig.~\ref{m1fermion:fig} we interpolated linearly between
Eq.~(\ref{m1FermionSmallc}) at $c = -0.48$ and
Eq.~(\ref{m1FermionNegc}) at $c = -0.5$.} We also show the exact
result (with the exact warp factor and scalar profile) obtained by
numerical integration (solid, red line), as well as the exact
numerical solution but assuming that $\phi(y)$ and $A(y)$ are
precisely given by Eqs.~(\ref{Approxphi}) and (\ref{ApproxA}) (solid,
blue line).  We restrict to the ``strong warping'' scenario, with $k/m
= 200$ (see Subsection~\ref{Linear}).  We see that the simple
expressions derived above are fairly reasonable whenever
$m_{1}/(k_{\rm eff}\,e^{-A(L)}) \lsim 1$.

For later use, we also give here the wavefunctions for the two
chiralities of the first (KK-parity odd) fermion KK mode.  The
left-handed chirality (or more generally, the chirality that coincides
with that of the zero-mode fermion in the tower) was found before in
terms of $\epsilon(y)$ of Eq.~(\ref{epsilonFermion}) on the interval
$[0,L]$.  Allowing for a normalization factor $N_{1}$, to be
determined from Eq.~(\ref{fermonNormalization}), we have for $y>0$:
\bea
f_L^1(y) &\approx& N_1 e^{(\frac{1}{2} - c)A(y)} \left[1 - m_1^2 \,
\int_y^L \! dz_2 \, e^{2A(z_{2})} \left( \frac{f^0(z_{2})}{f^0(0)} \right)^{-2}
\int_{z_2}^L \! dz_1 \left( \frac{f^0(z_{1})}{f^0(0)} \right)^2
\right]~,
\label{f1L}
\eea
where $f^{0}$ is the zero-mode wavefunction as given in
Eq.~(\ref{zeromodeSimple}).  For $y < 0$ one uses $f_L^1(-y) = -
f_L^1(y)$.  The wavefunction for the right-handed chirality (more
generally, the chirality opposite to that of the zero-mode in the
tower) is obtained from Eq.~(\ref{f_lFOeq}).  Using that
$e^{(\frac{1}{2} - c)A(y)}$ is the massless solution of
Eq.~(\ref{f_lFOeq}) [in the linear limit], $f_R^1$ is given on the
interval $[0,L]$ by
\bea
f_R^1(y) \approx  -N_1 m_1 \, e^{(\frac{1}{2} + c)A(y)} \int_y^L \! dz \,
\left( \frac{f^0(z)}{f^{0}(0)} \right)^{2}~,
\label{f1R}
\eea
which explicitly satisfies $f_R^1(L) = 0$.  For $y < 0$ one uses
$f_R^1(-y) = + f_R^1(y)$, i.e. that this wavefunction is even under $y
\rightarrow -y$.~\footnote{Since at $y=0$ we have $m_{D} = 0$ and $A'
= 0$, the fermion equations of motion (\ref{f_lFOeq}) and
(\ref{f_rFOeq}) imply that if one of the two chiralities has vanishing
first derivative at $y=0$ (an even mode) then the opposite chirality
vanishes at $y=0$ (an odd mode), and viceversa.} To the extent that
$m_{1} \ll k_{\rm eff}\,e^{-A(L)}$, we have that $\epsilon(y)$ is a
small effect and therefore the normalization constant $N_{1}$
coincides approximately with the normalization of the zero mode:
\bea
N_{1} &\approx& N_{0} ~\approx~  \sqrt{ \frac{2\left(1 - 2c \right) A(L)}{e^{(1 - 2c) A(L)} -1}}~,
\label{fermionN1N0}
\eea
where we used the approximate expression for the warp factor,
Eq.~(\ref{ApproxA}) [see Eq.~(\ref{lightfwavefunction}) below for more
details of the evaluation of $N_{0}$].

\section{The Higgs Field and EWSB}
\label{sec:Higgs}

The action for a (for simplicity, real) bulk scalar $H$ takes the form:
\bea
S_{\rm scalar} &= & \int \! d^{5}x \, \sqrt{g}
\left\{ \frac{1}{2} g^{MN} \partial_{M} H \partial_{N} H - V(H)
+ \delta(y+L) {\cal L}_{-L} + \delta(y-L) {\cal L}_{+L} \right\}~,
\label{BulkScalarAction}
\eea
where we allow for IR brane localized terms ${\cal L}_{\pm L}$ (these
$\delta$-terms should be written with the induced metric, but since in
our background we have $\sqrt{g} = \sqrt{g_{\rm ind}}$, we have
omitted this distinction for notational simplicity).  Focusing on the
quadratic terms, $V(H) = \frac{1}{2} M^{2} H^{2} + \cdots$, the KK
decomposition is written as
\bea
H(x^{\mu},y) = \frac{e^{A(y)}}{\sqrt{2L}} \sum^{\infty}_{n=0} H_{n}(x^{\mu}) f_{n}(y)~,
\label{KKScalar}
\eea
where the KK wavefunctions obey
\bea
f^{\prime\prime}_{n} - 2 A^{\prime} f^{\prime}_{n} +
\left[ A^{\prime\prime} - 3 A^{\prime 2} - M^{2} + e^{2A} m^{2}_{n} \right] f_{n} &=& 0~,
\label{ScalarEOM}
\eea
and are normalized according to
\bea \frac{1}{2L} \int_{-L}^L \! dy \, f_{n}(y) f_{m}(y)
&=& \delta_{nm}~.
\label{ScalarNormalization}
\eea
The boundary conditions associated with localized mass terms ${\cal
L}_{\pm L} = - \frac{1}{2} M_{\pm L} H^{2}$, where KK-parity imposes
$M_{- L} = M_{+ L}$, read
\bea
\left. \rule{0mm}{4mm} f^{\prime}_{n} + \left( A^{\prime} \pm M_{\pm L} \right) f_{n} \, \right|_{\pm L} &=& 0~.
\eea
It is convenient to parameterize the bulk and brane masses by
\bea
M^{2} &=& \left(\alpha^{2} - 4 \right) k^{2}_{\rm eff}~,
\hspace{1cm}
M_{\pm L} ~=~ \left(\alpha - 2 \right) k_{\rm eff}+ m_{\pm L} ~.
\label{Mdef}
\eea
where $k_{\rm eff} \equiv A^{\prime}(L)$, $\alpha \equiv c + 1/2$, and
$c = m_{D}(L)/A^{\prime}(L)$ as defined in Eq.~(\ref{cdefinition})
[see also Eqs.~(\ref{Approxphi}) and (\ref{ApproxA})].  In the
AdS$_{5}$ limit, where $A(y) = k y$, these equations admit a zero-mode
solution when $m_{\pm L} = 0$.  This follows by comparison with the KK
fermion Eq.~(\ref{fleq}), and the boundary condition (\ref{BCsfL}).
The zero-mode localization is controlled by the same $c$-parameter
that characterizes the fermion case.

We point out that for modes that are highly localized near the IR
boundaries, so that in the region $y > 0$ we can replace
$A^{\prime}(y) \approx A^{\prime}(L) = k_{\rm eff}$ and
$A^{\prime\prime}(y)/k^{2}_{\rm eff} \approx 1/(k_{\rm eff} L) \approx
1/[2A(L)] \ll 1$, Eq.~(\ref{ScalarEOM}) reduces to the fermion
Eq.~(\ref{fleq}) with $m_{D}(y) \approx m_{D}(L) = c A^{\prime}(L)$.
The scalar and fermion boundary conditions also coincide in this case,
up to the terms $m_{\pm L}$.  Thus, when $c \ll 1/2$ (strong fermion
zero-mode localization in the IR), the scalar EOM also has a light
mode, whose mass is controlled by $m_{\pm L}$.  To the extent that
$m_{\pm L}$ is small, the scalar and fermion spectra almost coincide
in this extreme IR localized limit.  In particular, from the same
considerations discussed in the radion and fermion cases of the
previous sections, there is a KK-parity odd mode that is exponentially
degenerate with the lightest scalar (KK-parity even) mode.

\subsection{Electroweak Symmetry Breaking}
\label{sec:EWSB}

A natural solution to the hierarchy problem suggests that the
Higgs field is localized near the IR boundaries.  In this case,
the Higgs KK modes become heavy and decouple, except for the
lightest KK-parity even and odd modes.  If the Higgs had a
vanishing VEV, the discussion of the previous section would imply
that these two light states are nearly degenerate, with their
masses controlled by the boundary mass terms $m_{\pm L}$
introduced above.  However, we need to introduce a potential that
leads to EWSB. It must also not break KK-parity, which implies
that only the KK-parity even Higgs mode can acquire a non-zero
VEV. The potential can arise from bulk or from IR localized terms,
but we may simply analyze the physics of EWSB in the low-energy 4D
effective theory.  The relevant 4D degrees of freedom are
KK-parity even and odd $SU(2)$ doublets with hypercharge $+1$,
that can be parameterized, respectively, as
\bea
H_{+} &=& \left(
\begin{array}{c}
G^{+}   \\
\tilde{v} + \frac{1}{\sqrt{2}} h_{+} + \frac{i}{\sqrt{2}} G^{0}
\end{array}
\right)~,
\hspace{1cm}
H_{-} ~=~ \left(
\begin{array}{c}
H^{+}   \\
\frac{1}{\sqrt{2}} h_{-} + \frac{i}{\sqrt{2}} a
\end{array}
\right)~,
\label{HiggsDoublets}
\eea
where $\tilde{v}$ denotes the warped-down VEV. Besides the kinetic
terms, $D_{\mu} H^{\dagger}_{+} D^{\mu}H_{+} + D_{\mu} H^{\dagger}_{-}
D^{\mu}H_{-}$, we can get potential terms of the form
\bea
V &=& m^{2}_{\rm odd} H^{\dagger}_{-} H_{-} + \frac{1}{2} \lambda_{1} ( H^{\dagger}_{-} H_{-})^{2} + \frac{1}{2} \lambda_{2} ( H^{\dagger}_{+} H_{+} -  \tilde{v}^{2})^{2} + \lambda_{3} ( H^{\dagger}_{-} H_{-}) ( H^{\dagger}_{+} H_{+} -  \tilde{v}^{2})
\nonumber \\ [0.5em]
&& \mbox{}
+ \lambda_{4} ( H^{\dagger}_{+} H_{-})( H^{\dagger}_{-} H_{+})
+ \frac{1}{2} \lambda_{5} \left[ ( H^{\dagger}_{+} H_{-})^{2}
+ ( H^{\dagger}_{-} H_{+})^{2} \right]~,
\label{EvenPotential}
\eea
where $m^{2}_{\rm odd}$ is the light KK mass discussed in the previous
subsection.  The above potential terms lead to masses $m^{2}_{h_{+}} =
2\lambda_{2} \tilde{v}^{2}$, $m^{2}_{h_{-}} = m^{2}_{\rm odd} +
(\lambda_{4} + \lambda_{5}) \tilde{v}^{2}$, $m^{2}_{a} = m^{2}_{\rm
odd} + (\lambda_{4} - \lambda_{5}) \tilde{v}^{2}$, and $m^{2}_{H^\pm}
= m^{2}_{\rm odd}$ ($G^{0}$ and $G^{\pm}$ are the eaten would-be
Goldstone bosons).

We note here, for later reference, that the KK-parity odd ``kinetic
terms'', $D_{\mu} H^{\dagger}_{+} D^{\mu}H_{-} + D_{\mu}
H^{\dagger}_{-} D^{\mu}H_{+}$, as well as the KK-parity odd potential
terms, $\lambda_{6} (H^{\dagger}_{+} H_{-} + H^{\dagger}_{-} H_{+})(
H^{\dagger}_{+} H_{+} + \zeta H^{\dagger}_{-} H_{-} - \tilde{v}^{2})$
can also appear, for instance when coupled to a KK-parity odd radion
excitation, $r'/\Lambda_{r}$.  For strongly IR localized Higgs fields
we expect $\lambda_{1} \approx \cdots \approx \lambda_{6}$ and $\zeta
\approx 1$.

\subsection{Yukawa Couplings to the Higgs and SM Fermion Masses}
\label{sec:Yukawas}

We consider next the fermion Higgs interactions.  Consider the top
Yukawa interaction
\bea
\delta S &=& - \int \! d^5x \, \sqrt{g}\;
\left[ \tilde{Y}_{5D} H \bar{Q} t + h.c. \right]~,
\label{HiggsYukawas}
\eea
where $\tilde{Y}_{5D}$ is the 5D Yukawa coupling (with mass dimension
$-1/2$), $H$ is the bulk Higgs doublet, $Q$ is the third generation
quark $SU(2)$ doublet, and $t$ is the top $SU(2)$ singlet [$Q$
satisfies the b.c.'s given in Eq.~(\ref{BCsfL}) and $t$ those given in
Eq.~(\ref{BCsfR})].

Focusing on the zero-mode fermions, as defined by
Eq.~(\ref{5Dfermion}), the effective 4D Yukawa coupling reads
\bea
Y_{4D} &=& \frac{\tilde{Y}_{5D}}{(2L)^{3/2}} \int_{-L}^{L} \! dy \,f_{h_{+}}(y) f^{0}_{L}(y)f^{0}_{R}(y)
~\approx~ \frac{Y_{5D}}{L} \, f^{0}_{L}(L)f^{0}_{R}(L)~,
\eea
where we took into account that the fermion zero-modes, as well as
$f_{h_{+}}$, are even about $y=0$, and the second equality holds in
the limit where the lightest Higgs state is strongly IR localized over
a distance of order $k_{\rm eff}$, so that $f_{h_{+}}(L) \approx
\sqrt{k_{\rm eff} L}$.  We also defined a 5D Yukawa coupling $Y_{5D} =
\tilde{Y}_{5D}/\sqrt{2k_{\rm eff}}$, as would be appropriate for a
localized Higgs field.  The general fermion zero-mode profiles are
given in Eq.~(\ref{zeromodes}), but here we use the approximations
given in Eq.~(\ref{zeromodeSimple}) with the warp factor given in
Eq.~(\ref{ApproxA}).  Recall that we use conventions such that for $c
> 1/2$ ($c < 1/2$) the zero-mode fermion is UV (IR) localized,
independently of chirality.  In addition, it is useful to write simple
analytical expressions for the properly normalized wavefunctions that
hold in the limit that $A(L) \gg 1$.  For $c > 1/2$ we use the fact
that for $z \gg 1$ one has $\int^z_{0} \!  dp \, e^{-p^2} \approx
\sqrt{\pi}/2$, so that Eq.~(\ref{zeromodeSimple}) reads
\bea
f^0(y) &\approx& \left[ \frac{8\left(c-\frac{1}{2}\right) A(L)}{\pi} \right]^{1/4} e^{-(c-\frac{1}{2}) A(y)}~,
\hspace{1cm}
{\rm for }~c > 1/2~.
\eea
For $c < 1/2$ one has instead $\int^z_{0} \!  dp \, e^{p^2} \approx
\frac{1}{2z} \, e^{z^2}$ whenever $z \gg 1$, and therefore
\bea
f^0(y) &\approx& \sqrt{ \frac{4 \left(\frac{1}{2} - c \right) A(L)}{e^{(1 - 2c)
A(L)}}} \, e^{(\frac{1}{2} - c) A(y)}~,
\hspace{1cm}
{\rm for }~c < 1/2~.
\label{lightfwavefunction}
\eea
These approximations work extremely well except very close to $c =
1/2$, where the wavefunction is (approximately) flat: $f^0(y) \approx
1$.  Then, for instance, if the LH top has a flat profile ($c_{L} =
1/2$), while the RH top is localized near the IR boundaries ($c_{R} <
1/2$) the 4D top Yukawa coupling reads
\bea
Y_{4D} &\approx& \sqrt{\rule{0mm}{3.5mm} \left(1 - 2c_{R} \right) k_{\rm eff} L} \times \frac{Y_{5D}}{L}~,
\eea
where we used Eq.~(\ref{ApproxA}) for $A(L)$.

We estimate the maximum value of $Y_{5D}$ from Naive Dimensional
Analysis (NDA) in extra dimensions with
singularities~\cite{Chacko:1999hg}.  Writing the 5D Yukawa coupling as
$Y_{5D} = w/\Lambda_{5}$, where $\Lambda_{5}$ is the cutoff scale of
the 5D theory, NDA gives $w_{\rm NDA} \sim l_5/\sqrt{l_4}$, where $l_5
= 24\pi^3$ and $l_4 = 16 \pi^2$ are the 5D and 4D loop factors,
respectively.  Also, taking $\Lambda_{5}$ as the scale where 5D QCD
gets strong gives
\begin{equation}
\Lambda_{5} L\sim l_5/N_c\; ,
\end{equation}
where $N_c = 3$ is the number of colors. Therefore, we
estimate
\bea
\frac{Y^{\rm NDA}_{5D}}{L} &\sim& \frac{N_{c}}{\sqrt{l_4}} ~=~ \frac{3}{4\pi}~.
\label{Y5DNDA}
\eea
\begin{figure}
\centerline{ \hspace*{-0.5cm}
\includegraphics[width=0.47 \textwidth]{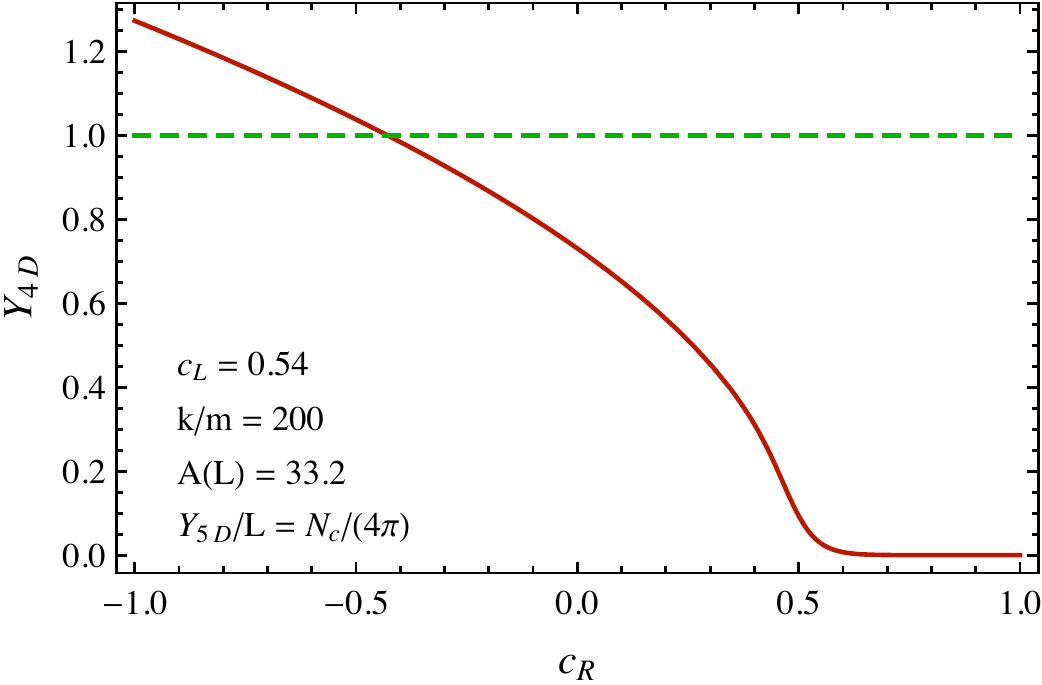}
\hspace*{0.5cm}
\includegraphics[width=0.47 \textwidth]{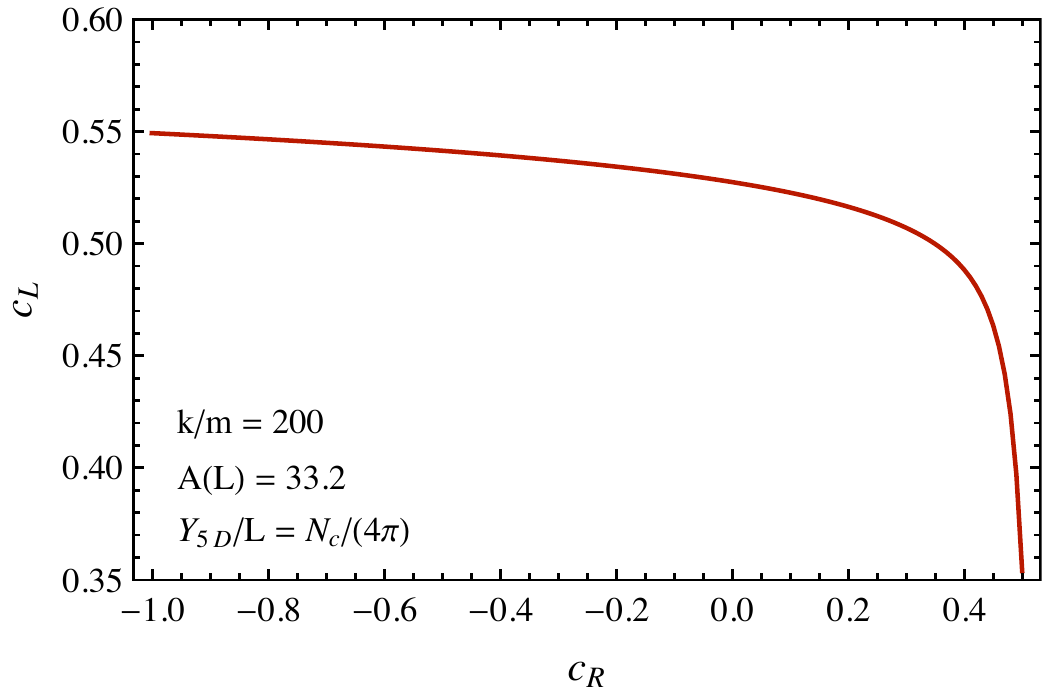}
} \caption{\em Left panel: Four-dimensional Yukawa coupling as a
function of $c_{R}$, for fixed $c_{L} =0.54$, $k/m=200$ and $A(L)
= 33.2$.  The dashed (green) line shows the approximate value of
the top Yukawa coupling which occurs for $c_{R}\sim -0.45$.  Right
panel: $c_{L}$ as a function of $c_{R}$ for the observed top mass
($Y_{4D} \approx 1$), assuming $Y_{5D} = N_{c}/(4\pi)$, with
$N_{c} = 3$.} \label{topYukawa:fig}
\end{figure}
In the left panel of Fig.~\ref{topYukawa:fig} we show the top Yukawa
coupling as a function of $c_{R}$ for $c_{L} = 0.54$, taking $Y_{5D}$
as given in Eq.~(\ref{Y5DNDA}) [and in the ``strong warping''
benchmark scenario of Subsection~\ref{Linear}].  We use the exactly
normalized zero-mode fermion profiles, although the approximate
formulas given above lead to very accurate results except in the close
vicinity of $c=1/2$.  Notice that we can generate a 4D Yukawa coupling
of order one with a flat --or even slightly localized on the UV brane
($c_L\sim 0.54$)-- left-handed zero mode, and a right-handed zero mode
near the IR brane ($c_{R}\sim -0.45$).  For these values,
Fig.~\ref{m1fermion:fig} shows that there is a KK-parity odd resonance
of the $SU(2)$ singlet top with a mass $m_{1}\sim 0.44 \,
\tilde{k}_{\rm eff}$.

In the right panel of Fig.~\ref{topYukawa:fig}, we show the required
$c_{L}$ as a function of $c_{R}$, that reproduces the observed 4D top
Yukawa coupling, again assuming that the 5D Yukawa coupling takes the
NDA value of Eq.~(\ref{Y5DNDA}).  We see that the $SU(2)$ top doublet
has $c_{L} \sim 0.5$, except when $c_{R}$ approaches $1/2$.

\section{Gauge Fields}
\label{sec:gauge}

For completeness, we summarize here the KK decomposition for gauge
fields (in the absence of brane kinetic terms).  The results apply for
general diagonal metrics of the form~(\ref{metric}).  The gauge action
reads
\bea
S &=& \int \! d^{5}x \, \sqrt{g} \left\{
- \frac{1}{4} g^{MK} g^{NL} F_{MN} F_{KL} \right\} + S_{\rm g.f.}~.
\label{Sgauge}
\eea
The gauge fixing term in the background of Eq.~(\ref{metric}) is
chosen as~\cite{Randall:2001gb}
\bea
S_{\rm g.f.} &=& - \frac{1}{2\xi} \int \! d^{5}x \,
\left\{ \partial_{\mu} V^{\mu} - \xi \, \partial_{y} \left[ e^{-2A} V_{5}
\right] \right\}^{2}~,
\eea
where $\xi$ is the gauge fixing parameter, and the 4D indices are
contracted with $ \eta_{\mu\nu}$.
The KK mode expansions then read
\bea
V_{\mu,5}(x,y) &=& \frac{1}{\sqrt{2L}} \sum_{n=0}^{\infty} V_{\mu,5}^n(x) f_{V,5}^n(y)~,
\label{GaugeKK}
\eea
where
\bea
\partial_{y} \left[ e^{-2A} \partial_{y} f_{V}^{n} \right] + m_{n}^{2} f_{V}^{n} &=& 0~,
\label{GaugeEOM}
\eea
and $f_{5}^{n} = \partial_{y} f_{V}^{n}/m_{n}$.  For SM gauge fields,
the boundary conditions are taken as $\partial_{y} f^{n}_{V} |_{0,L} =
f^{n}_{5} |_{0,L} = 0$, so that there is no zero-mode for $V_{5}$.  In
the above $R_{\xi}$-type of gauge, $V_{5}^{n}$ has mass $\xi
m^{2}_{n}$.  The KK wavefunctions obey the orthonormality relation
\bea
\frac{1}{2L}\int_{-L}^L \! dy\, f_V^n f_V^m &=& \delta_{nm}~.
\label{GaugeNormalization}
\eea
The gauge eigenvalues in the ``strong warping'' benchmark scenario
defined in Subsection~\ref{Linear} are approximately given in
Fig.~\ref{fig:roots}. The gauge zero-mode wavefunction, however, is
always exactly flat ($f^{0}_{V} = 1$), unlike the zero-mode
wavefunction for a fermion with $c = 1/2$.

\section{Radion and KK-Radion Couplings at Linear Order}
\label{sec:Rcouplings}

In previous sections we saw that, apart from the ``zero-mode'' sector
(that includes the SM fields plus the radion), some first-level KK
states can be relatively light compared to the KK scale.  These light
states are KK-parity odd.  In particular, in the present scenario the
lightest KK-parity odd particle (the LKP) is the first level KK-radion
while the second lightest KK state (the NLKP) corresponds to either
the first KK-parity odd mode of the IR localized top $SU(2)$ singlet,
or the first KK-parity odd excitation of the Higgs doublet.  In this
section, we work out the Feynman rules for the interactions involving
a radion/KK-radion with the top KK tower, and give some numerical
examples.  We also give the radion/KK-radion couplings to the gauge KK
towers, and also to the two Higgs doublets.

\subsection{KK-Radion Couplings to Fermions}
\label{sec:Rffcouplings}

The couplings of the radion KK tower to the KK fermions are obtained
by replacing into the action the KK decompositions for the metric and
bulk scalar, as given in
Eqs.~(\ref{MetricFluctuations})-(\ref{RadionKKDecomposition}),
together with the fermion KK decomposition given in
Eq.~(\ref{5Dfermion}).  We are interested in the terms linear in the
radion or its first KK-parity odd excitation.~\footnote{Note that the
couplings involving heavier radion KK states receive a further
suppression from the $1/m_{n}^{2}$ in the radion normalization,
Eq.~(\ref{RadionOrthonormality}) [in other words, heavier radion modes
have an effectively larger decay constant, $\Lambda^{n}_{r} \sim x_{n}
\Lambda_{r}$, where $m_{n} = x_{n} \, k_{\rm eff} \, e^{-A(L)}$ is the
$n$-th KK radion mass].} There are two types of contributions.  First,
those arising from the bulk fermion action, Eq.~(\ref{FermionAction}),
which involve KK fermions belonging to the same 5D fermion field.
Second, those arising from the localized Yukawa interactions, as in
Eq.~(\ref{HiggsYukawas}), which involve KK fermions belonging to
different 5D fermion fields.

The Yukawa interactions lead to mass mixing (after EWSB) between
different fermion KK towers, most importantly for the top tower.
Although these effects can be taken into account perturbatively, it is
safer to perform a KK decomposition that already incorporates EWSB
effects (for an IR-localized Higgs field, the bulk solutions worked
out before do not change, but the boundary conditions are deformed by
EWSB).  In this way, the effects of EWSB appear directly in the
vertices and spectrum, and one does not have to worry about mixing in
external legs of physical processes.  A second possibility is to
diagonalize the fermion mass matrix including a sufficiently large
number of KK modes.  Here, for simplicity, we will derive the Feynman
rules for the vertices \textit{ignoring} EWSB. This should contain the
essential physics for most applications (but see
Ref.~\cite{Medina:2011qc} for a discussion of EWSB effects).  However,
note that the localized Yukawa terms also lead to quartic interactions
involving the physical Higgs, $h$, of the form $h{\rm -}r_{i}{\rm
-}t^{j}{\rm -}t^{k}$ that are proportional to the EWSB zero-mode
fermion mass, $m_{t}$.

As just stated, we proceed now by setting EWSB to zero.  In terms of
the canonically normalized radion KK modes, $r_{n}(x) \equiv
\Lambda_{r} \tilde{r}_{n}(x)$, where $\Lambda_{r}$ is the radion decay
constant defined in Eq.~(\ref{rdecayconst}), the interactions induced
by the bulk action take the form
\bea
\mathcal{L}_{\rm r\psi \psi}^{\rm bulk} &=& - \sum_{ijk=0}^\infty \, \frac{r_i}{\Lambda_{r}} \left[
g_{ijk}^{LL} \, \bar{\psi}_L^{j} \, i \! \stackrel{\leftrightarrow}{\slash{\!\!\!\partial}} \! \psi_L^k +
g_{ijk}^{RR} \, \bar{\psi}_R^{j} \, i \! \stackrel{\leftrightarrow}{\slash{\!\!\!\partial}} \! \psi_R^k \right]
+ \sum_{ijk=0}^\infty \, \frac{r_i}{\Lambda_{r}} \left[ m^{RL}_{ijk} \,
\bar{\psi}_R^{j} \psi_L^k + \mbox{h.c.} \right]~,
\hspace{3mm}
\label{Lrffbulk}
\eea
where $\bar{\psi} \!
\stackrel{\leftrightarrow}{\slash{\!\!\!\partial}} \!\!  \chi \equiv
\frac{1}{2} \left[ \bar{\psi} \gamma^{\mu} \partial_{\mu} \chi -
(\partial_{\mu} \bar{\psi}) \gamma^{\mu} \chi \right]$,
\bea
g^{LL}_{ijk} &=& \frac{1}{2L} \int_{-L}^L \! dy \, F_i f_L^j f_L^k~,
\hspace{1.5cm}
g^{RR}_{ijk} ~=~ \frac{1}{2L} \int_{-L}^L \! dy \, F_i f_R^j f_R^k~,
\label{grLL}\\
m^{RL}_{ijk} &=& 2 g^{LL}_{ijk} \, m_{j} + 2 g^{RR}_{ijk} \, m_{k} - 
\frac{1}{2L} \int_{-L}^L dy \, e^{-A}\left[2 m_D F_i
+ \frac{m_D'}{A''} e^{2A} (e^{-2A} F_i)' \right]  f_R^j f_L^k~,
\label{mijk}
\eea
and $m^{LR}_{ijk} = m^{RL}_{ikj}$.  Here we have used the fact that
the f\"unfbein is diagonal with $e^\mu_{\nu} \approx e^{A}(1+F) \,
\delta^\mu_{\nu}$ and $e^5_{5} \approx (1 - 2F)$.  To arrive at
Eq.~(\ref{mijk}) we used the fermion equations of motion,
Eq.~(\ref{f_lFOeq}) and (\ref{f_rFOeq}).  The second term in the
integral in Eq.~(\ref{mijk}) arises from the scalar fluctuations given
in Eq.~(\ref{RadionScalarConnection}), where we used the background
relation $\phi^{\prime 2} = 3 M_{5}^{3} A''$ to eliminate $M_{5}$.
Note also that the integrands in Eq.~(\ref{grLL}) and Eq.~(\ref{mijk})
are odd when $(i+j+k) \neq 0 \; {\rm mod} \; 2$, which implies
$g_{ijk}^{LL} = g_{ijk}^{RR} = m^{RL}_{ijk} = 0$ in such cases, so
that KK-parity is not violated.

It is of some interest to consider the interactions involving $n$
radion modes and a fermion pair.  The $r_{i_{1}} \cdots r_{i_{n}}
\bar{\psi}^{j} \psi^{k}$ interactions take a similar form to
Eq.~(\ref{Lrffbulk}) with $\Lambda_{r} \to \Lambda_{r}^{n}$,
$g^{LL}_{ijk} \to c_{n} \, g^{LL}_{i_{1} \ldots i_{n}jk}$,
$g^{RR}_{ijk} \to c_{n} \, g^{RR}_{i_{1} \ldots i_{n}jk}$ and
$m^{RL}_{ijk} \to d_{n} \, m^{RL}_{i_{1} \ldots i_{n}jk}$, where
$c_{n} = (-3)^{n} \left( 2n/3 - 1 \right)/n!$ and $d_{n} =
(-4)^{n-1}/n!$ are combinatorial factors, and
\bea
g^{LL}_{i_{1} \ldots i_{n}jk} &=& \frac{1}{2L} \int_{-L}^L \! dy \, F_{i_{1}} \cdots F_{i_{n}} f_L^j f_L^k~,
\hspace{8mm}
g^{RR}_{i_{1} \ldots i_{n}jk} ~=~ \frac{1}{2L} \int_{-L}^L \! dy \, F_{i_{1}} \cdots F_{i_{n}}  f_R^j f_R^k~,
\label{grnLL}\\
m^{RL}_{i_{1} \ldots i_{n}jk} &=& 2 g^{LL}_{i_{1} \ldots i_{n}jk} \, m_{j}
+ 2 g^{RR}_{i_{1} \ldots i_{n}jk} \, m_{k}
- \frac{1}{L} \left( 2 - n \right) \int_{-L}^L dy \, e^{-A} m_D F_{i_{1}} \cdots F_{i_{n}} f_R^j f_L^k
\nonumber \\[0.5em]
& & \mbox{} - \frac{1}{2L} \frac{n}{2} \left( 3 - n \right) \int_{-L}^L dy \, \frac{m_D'}{A''} e^{A} F_{i_{1}} \cdots F_{i_{n-1}} (e^{-2A} F_{i_{n}})' f_R^j f_L^k~.
\label{mi1injk}
\eea
It is understood that in the Lagrangian we write separate terms for
each $(i_{1}, \ldots, i_{n})$ ordered $n$-tuple.  For instance, in the
case $n = 2$, if $i_{1} \neq i_{2}$ we have two terms corresponding to
$r_{i_{1}} r_{i_{2}} \bar{\psi}^{j} \psi^{k}$ and $r_{i_{2}} r_{i_{1}}
\bar{\psi}^{j} \psi^{k}$, with the coefficients given above.  If
$i_{1} = i_{2}$ then there is a single term.  This case plays a role
in the annihilation processes $r_{i} r_{i} \to \bar{\psi} \psi$.

The bulk action also leads to interactions involving the radion,
fermion pairs and gauge bosons $V_{\mu}$ of the form
\bea
\mathcal{L}_{r\psi\psi V} = - \sum_{ijlk} \frac{r_i}{\Lambda_{r}} \left( g_{ijlk}^{LL} \, \bar{\psi}_L^j
\slash{\!\!\!V}^l \psi_L^k + g_{ijlk}^{RR} \, \bar{\psi}_R^j \slash{\!\!\!V}^l
\psi_R^k \right)~,
\label{LrffV}
\eea
where we used the KK expansion for the gauge fields,
Eq.~(\ref{GaugeKK}), and defined
\bea
g_{ijlk}^{LL} &=& \frac{g_V}{2L} \, \int_{-L}^L \! dy\,
F_i f_L^j f_V^l f_L^k \label{grVLL}~,
\\
g_{ijlk}^{RR} &=& \frac{g_V}{2L} \, \int_{-L}^L \! dy\,
F_i f_R^j f_V^l f_R^k~,
\label{grVRR}
\eea
with $g_V = g_{5\,V}/\sqrt{2L}$ the 4D gauge coupling.  We work in
unitary gauge so that there is no $\bar{\psi}_L^j V_{5 }\psi_R^k$
interaction (the $V_5$ components of the gauge field are eaten by the
higher KK modes, and the zero-mode of $V_5$ is projected out by the
boundary conditions).  Note that when we consider the interactions
involving a zero-mode gauge boson, with $f_V^0(y) = 1$ up to EWSB
effects, we have $g_{ijlk}^{LL} = g_{V} g_{ijk}^{LL}$ and
$g_{ijlk}^{RR} = g_{V} g_{ijk}^{RR}$, where $g_{ijk}^{LL}$ and
$g_{ijk}^{RR}$ are defined in Eq.~(\ref{grLL}).

The Feynman rules for a single radion state, $r_{i}$, and two KK
fermions or a pair of KK fermions and a (neutral) gauge boson are then
\bea
\put(-30,-32){
\resizebox{3cm}{!}{
\includegraphics{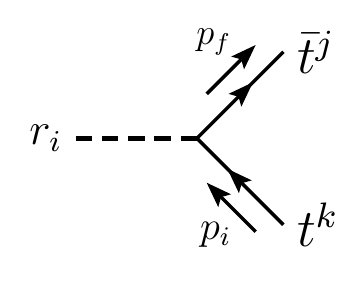}}
}
\hspace*{2cm}
&=&
- \frac{i}{\Lambda_r}\left\{\frac{1}{2}(\slash{\!\!\!p}_i+\slash{\!\!\!p}_f) \left( g^{LL}_{ijk} P_L + g^{RR}_{ijk} P_R \right) - m^{RL}_{ijk} P_L - m^{LR}_{ijk} P_R  \right\}~,
\nonumber
\\ [0.5em]
\rule{2cm}{0mm}
\put(-30,-38){
\resizebox{3cm}{!}{
\includegraphics{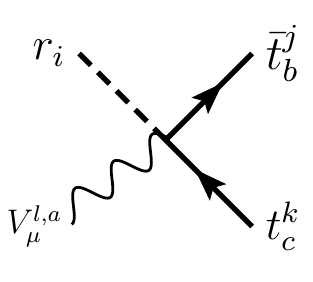}}
}
\hspace*{2cm}
&=& -\frac{i}{\Lambda_r} \gamma^{\mu} (T^{a})_{bc} \left( g^{LL}_{ijlk} P_L + g^{RR}_{ijlk} P_R \right)~,
\nonumber
\eea
which, as mentioned above, are non-vanishing only when $(i+j+k) =
0 \; {\rm mod} \; 2$, or $(i+j+k+l) = 0 \; {\rm mod} \; 2$.  For
the Feynman rule involving the gauge boson: in the case of the
photon, $(T^{a})_{bc} \rightarrow 1$ and $g_{V} = Q_{t} e$, with
$Q_{t} = \frac{2}{3}$ the top quark electric charge; for the $Z$
gauge boson, $(T^{a})_{bc} \rightarrow 1$ and $g_{V} = \sqrt{g^2 +
g^{\prime 2}} \, (T^{3}_{t} - Q_{t} s_{W}^2)$; for the gluons,
$g_{V} = \sqrt{4\pi\alpha_{s}}$, with $T^{a}$ the $SU(3)$
generators normalized according to ${\rm Tr}(T^{a}T^{b}) =
\frac{1}{2} \delta^{ab}$, and $a, b, c$ are color indices.

The vertices involving the lightest states (the radion, the first KK
radion, the RH top and its first KK excitation) are especially
important, and their Feynman rules can be further simplified.  We can
also restrict to couplings involving the zero-mode gauge bosons.
Since the radion, $r$ [i.e. the lightest of the scalar fluctuations in
the 5D metric/bulk scalar system, with a mass $m_{0}$ as given in
Eqs.~(\ref{mRadionExact})-(\ref{mRadionApprox2})], is KK-parity even,
it can only couple to a pair of KK-parity even or a pair of KK-parity
odd fermions.  Similarly, the first radion excitation [which we call
$r'$ and is essentially degenerate with $r$ (see
Section~\ref{RadionOddScalar})], is KK-parity odd and can couple to
$t$ and $t'$, where we denote by $t'$ the first KK excitation of the
SU(2) singlet top.  The couplings involving these light fields can be
approximated based on the following observations:

\begin{enumerate}

\item The radion wavefunction is highly peaked near the IR boundaries
over a distance of order $1/k_{\rm eff}$, as given in
Eq.~(\ref{F0Norm}).  Also, as discussed in
Section~\ref{sec:LightFermion}, $r'$ has a wavefunction essentially
identical to that of the radion: $F_{1}(y) \approx F_{0}(y)$.

\item For the RH top, $f_R^1(y) \approx f_R^0(y)$ on the interval
$[0,L]$, and both the zero-mode and its first KK excitation, $t'$, are
highly peaked near the IR boundaries.  Recall that $f_R^0(y)$ was
given in Eq.~(\ref{lightfwavefunction}).

\item The wavefunction for the LH chirality of $t'$, $f_L^1(y)$ [given
in Eq.~(\ref{f1R}), with the trivial change in label $R \rightarrow
L$] is more spread-out over the full extent of the extra-dimension.
To see this, note that the IR localization of $f_R^0(y)$ implies that
the integral in Eq.~(\ref{f1R}) is almost y-independent, except very
close to $y = L$, where it vanishes.  Hence the main source of
$y$-dependence arises from the $e^{(\frac{1}{2} + c_{R}) A(y)}$
factor.  For $c_{R} \sim -1/2$ this factor is also nearly constant, so
that the normalization implies that $f_L^1(y) \approx 1$.

\end{enumerate}

It follows that for $c_{R} \sim -1/2$ the couplings involving
$f_R^0(y)$ or $f_R^1(y)$ are larger than those involving $f_L^1(y)$ by
a factor of order $f_R^0(L) \sim \sqrt{2(1 - 2c_{R}) A(L)} \sim
\sqrt{k_{\rm eff}L}$, for each occurrence of the RH versus LH
wavefunction, so that for instance $g^{LL}_{101} \sim
g^{RR}_{101}/(k_{\rm eff} L)$ in Eq.~(\ref{grLL}).  When $c_{R}$ is
not so close to $-1/2$ the suppression is less severe, but we still
have $g^{LL}_{101} \ll g^{RR}_{101}$ provided $f_R^0(y)$ is IR
localized.  The integral in Eq.~(\ref{mijk}) is similarly suppressed.
In fact, the second term inside the integral goes like $e^{-A} g'
F_{0} f^{0}_{R} f^{1}_{L}$, where $g'$, as given in
Eq.~(\ref{g'wavefunction}), gives an additional suppression that makes
this contribution arising from the scalar fluctuations completely
negligible.  When focusing on the couplings to the lightest top KK
state, we can therefore neglect all the contributions that depend on
$f_L^1(y)$.  For typical choices of parameters the errors thus induced
are of order $(10{\rm -}20) \%$.

The upshot is that the bulk interactions between the radion/LKP, the
NLKP (with both RH and LH chiralities) and the top quark, as well as
those involving a SM gauge boson, are controlled by
\bea
g^{RR}_{000} &\approx& g^{RR}_{011} ~\approx~ g^{RR}_{101} ~=~
g^{RR}_{110} ~\approx~ \left(\frac{1 - 2c}{3-2c}\right) s_{A}^{1/2}~.
\label{g000}
\eea
For the radion couplings to a fermion pair, the explicit Feynman rules
are:
\bea
\put(-30,-27){
\resizebox{3cm}{!}{
\includegraphics{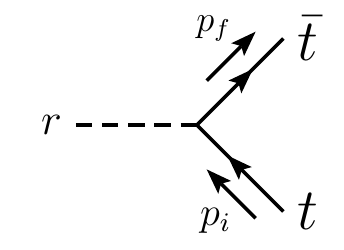}}
}
\hspace*{2cm}
 &\approx& -i \frac{g^{RR}_{000}}{\Lambda_r} \, \frac{1}{2}
 \left( \slash{\!\!\!p_{i}} + \slash{\!\!\!p_{f}} \right)~,
\\ [0.5em]
\put(-30,-27){
\resizebox{3cm}{!}{
\includegraphics{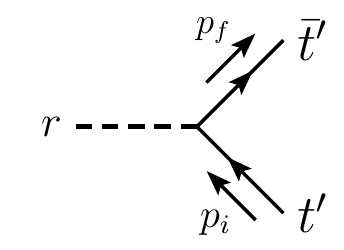}}
}
\hspace*{2cm}
&\approx&
- i \frac{g^{RR}_{011}}{\Lambda_r}\left[ \frac{1}{2}
\left( \slash{\!\!\!p_{i}} + \slash{\!\!\!p_{f}} \right) P_{R} - 2 m_{t^\prime} \right]~.
\eea
The mass of the first KK resonance, $m_{t^{\prime}}$, is given in
Eq.~(\ref{m1FermionSmallc}).  Similarly, the couplings involving $r'$
are:
\bea
\put(-30,-27){
\resizebox{3cm}{!}{
\includegraphics{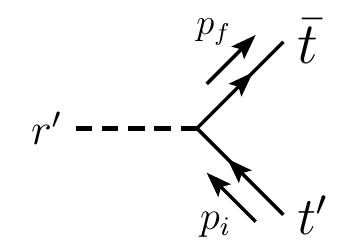}}
}
\hspace*{2cm}
&\approx&
- i \frac{g^{RR}_{101}}{\Lambda_r}\left[ \frac{1}{2}
\left( \slash{\!\!\!p_{i}} + \slash{\!\!\!p_{f}} \right) P_{R} - 2 m_{t^\prime} P_{L} \right]~,
\label{rpttp}
\\ [0.5em]
\rule{2cm}{0mm}
\put(-30,-38){
\resizebox{3cm}{!}{
\includegraphics{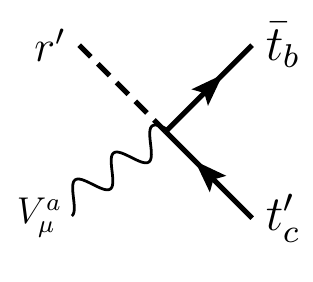}}
}
\hspace*{2cm}
&\approx& -i \frac{g^{RR}_{101}}{\Lambda_r} g_{V} \gamma^{\mu} (T^{a})_{bc} P_{R}~,
\label{Vrpttp}
\eea
where $V_{\mu}$ stands for any of the neutral SM gauge bosons.  We did
not write explicitly the hermitian conjugate processes.  The rules for
the $V_{\mu}{\rm -}r{\rm -}t{\rm -}\bar{t}$ and $V_{\mu}{\rm -}r{\rm
-}t'{\rm -}\bar{t}'$ vertices are analogous to Eq.~(\ref{Vrpttp}),
with $g^{RR}_{101} \to g^{RR}_{000}$ and $g^{RR}_{101} \to
g^{RR}_{011}$, respectively.

For illustration, we give an example of the relevant couplings for the
radion [$g^{LL}_{0jk}$, $g^{RR}_{0jk}$ and $m^{LR}_{0jk}/k_{\rm
eff}\,e^{-A(L)}$] and for the KK-radion [$g^{LL}_{1jk}$,
$g^{RR}_{1jk}$ and $m^{LR}_{1jk}/k_{\rm eff}\,e^{-A(L)}$], to the
lowest lying KK fermions.  In Table~\ref{radionCoupStrongWarp} we give
the couplings in the ``strong warping'' benchmark scenario, while
Table~\ref{radionCoupSmallWarp} contains the corresponding couplings
in the ``small warping'' benchmark scenario (these scenarios are
defined in Subsection~\ref{Linear}).  We chose $c = -0.2$, as might be
appropriate for the $SU(2)$ top singlet.  In this case, the fermion
masses are given by $m_{n} = x^{f}_{n} \, e^{-A(L)}$, with $x^{f}_{n}
= 0, 1.15, 3.38, 4.32, 6.50, 7.43, \ldots$ ($x^{f}_{n} = 0, 0.97,
2.96, 3.78, 5.63, 6.51, \ldots$) in the strong (small) warping
benchmark scenarios.  Even (odd) $n$ correspond to KK-parity even
(odd) states.  For such $c$-values, the KK-parity even and odd states
are not as degenerate as for $c = 1/2$, but one can recognize the
corresponding semi-degenerate pairs.  Note that the tables display the
selection rules implied by KK-parity conservation.  We also point out
that the last integral in the definition of $m^{RL}_{ijk}$ [see
Eq.~(\ref{mijk})] typically gives a small contribution, although it
has been included in the numerical evaluation of
Tables~\ref{radionCoupStrongWarp} and \ref{radionCoupSmallWarp}.  We
see in the tables that the couplings involving fermions other than the
zero-mode and/or the first (``ultra-light'') fermion excitation, are
almost vector-like (see the $4\times 4$ submatrices in the lower right
corner of each table).  The couplings involving $t$ and $t'$, on the
other hand, are close to chiral, as we have also exhibited in
Eqs.~(\ref{rpttp}) and (\ref{Vrpttp}).

\begin{table}[t]
\vspace*{-2cm}
\begin{center}
\begin{tabular}{|c||cccccc|}
\multicolumn{7}{l}{$\bf g^{LL}_{0jk}$} \\ [0.4em]
\hline
\rule{0mm}{4mm}
$x^{f}_{n}$  & \bf 0  & \bf 1.15  & \bf 3.38  & \bf 4.32  & \bf 6.50  & \bf 7.43  \\ [0.2em]
\hline
\hline
\bf 0            &  0      &  0      &  0       &  0      &  0      & 0       \\
\bf 1.15       &  0      &  0.08 &  0      &  0.13 &  0      & 0.02  \\
\bf 3.38       &  0      &  0      &  0.30 &  0      &  0.18 & 0       \\
\bf 4.32       &  0      &  0.13 &  0      &  0.31 &  0      & 0.19  \\
\bf 6.50       &  0      &  0      &  0.18 &  0      &  0.32 & 0       \\
\bf 7.43       &  0      &  0.02 &  0      &  0.19 &  0      & 0.32  \\
\hline
\end{tabular}
%
\begin{tabular}{|cccccc|}
\multicolumn{6}{l}{$\bf g^{LL}_{1jk}$} \\ [0.4em]
\hline
\rule{0mm}{4mm}
\bf 0  & \bf 1.15  & \bf 3.38  & \bf 4.32  & \bf 6.50  & \bf 7.43  \\ [0.2em]
\hline
\hline
0      &  0      &  0      &  0      &  0      & 0       \\
0      &  0      &  0.15 &  0      &  0.05 & 0       \\
0      &  0.15 &  0      &  0.29 &  0      & 0.12  \\
0      &  0      &  0.29 &  0      &  0.24 & 0       \\
0      &  0.05 &  0      &  0.24 &  0      & 0.31  \\
0      &  0      &  0.12 &  0      &  0.31 & 0       \\
\hline
\end{tabular}
\end{center}
\begin{center}
\begin{tabular}{|c||cccccc|}
\multicolumn{7}{l}{$\bf g^{RR}_{0jk}$} \\ [0.4em]
\hline
\rule{0mm}{4mm}
$x^{f}_{n}$  & \bf 0  & \bf 1.15  & \bf 3.38  & \bf 4.32  & \bf 6.50  & \bf 7.43  \\ [0.2em]
\hline
\hline
\bf 0            &  0.41 &  0      &  0.29 &  0      &  0.08 & 0       \\
\bf 1.15       &  0      &  0.52 &  0      &  0.26 &  0      & 0.08  \\
\bf 3.38       &  0.29 &  0      &  0.35 &  0      &  0.22 & 0       \\
\bf 4.32       &  0      &  0.26 &  0      &  0.34 &  0      & 0.21  \\
\bf 6.50       &  0.08 &  0      &  0.22 &  0      &  0.34 & 0       \\
\bf 7.43       &  0      &  0.08 &  0      &  0.21 &  0      & 0.33  \\
\hline
\end{tabular}
%
\begin{tabular}{|cccccc|}
\multicolumn{6}{l}{$\bf g^{RR}_{1jk}$} \\ [0.4em]
\hline
\rule{0mm}{4mm}
\bf 0  & \bf 1.15  & \bf 3.38  & \bf 4.32  & \bf 6.50  & \bf 7.43  \\ [0.2em]
\hline
\hline
0      &  0.46 &  0      &  0.20 &  0      & 0.06  \\
0.46 &  0      &  0.35 &  0      &  0.11 & 0       \\
0      &  0.35 &  0      &  0.33 &  0      & 0.16  \\
0.20 &  0      &  0.33 &  0      &  0.27 & 0       \\
0      &  0.11 &  0      &  0.27 &  0      & 0.32  \\
0.06 &  0      &  0.16 &  0      &  0.32 & 0       \\
\hline
\end{tabular}
\end{center}
\begin{center}
\begin{tabular}{|c||cccccc|}
\multicolumn{7}{l}{$\bf m^{LR}_{0jk}/\tilde{k}_{\rm eff}$} \\ [0.4em]
\hline
\rule{0mm}{4mm}
$x^{f}_{n}$  & \bf 0  & \bf 1.15  & \bf 3.38  & \bf 4.32  & \bf 6.50  & \bf 7.43  \\ [0.2em]
\hline
\hline
\bf 0            &  0      &  0      &  0      &  0      &  0      & 0       \\
\bf 1.15       &  0      &  1.45 &  0      &  1.68 &  0      & 0.52  \\
\bf 3.38       &  2.05 &  0      &  4.43 &  0      &  3.78 & 0       \\
\bf 4.32       &  0      &  2.67 &  0      &  5.67 &  0      & 4.57  \\
\bf 6.50       &  1.09 &  0      &  4.18 &  0      &  8.56 & 0       \\
\bf 7.43       &  0      &  1.27 &  0      &  4.90 &  0      & 9.79  \\
\hline
\end{tabular}
%
\begin{tabular}{|cccccc|}
\multicolumn{6}{l}{$\bf m^{LR}_{1jk}/\tilde{k}_{\rm eff}$} \\ [0.4em]
\hline
\rule{0mm}{4mm}
\bf 0  & \bf 1.15  & \bf 3.38  & \bf 4.32  & \bf 6.50  & \bf 7.43  \\ [0.2em]
\hline
\hline
0      &  0      &  0      &  0      &  0      & 0       \\
1.12 &  0      &  1.78 &  0      &  0.86 & 0       \\
0      &  2.80 &  0      &  4.77 &  0      & 2.89  \\
1.88 &  0      &  4.91 &  0      &  5.45 & 0       \\
0      &  1.64 &  0      &  5.73 &  0      & 8.74  \\
0.87 &  0      &  3.33 &  0      &  8.83 & 0       \\
\hline
\end{tabular}
\end{center}
\begin{center}
\begin{tabular}{|c||ccccc|}
\multicolumn{6}{l}{$\bf g_{1jk}$} \\ [0.4em]
\hline
\rule{0mm}{4mm}
$x^{V}_{n}$  & \bf 0  & \bf 2.39  & \bf 2.41  & \bf 5.46  & \bf 5.49  \\ [0.2em]
\hline
\hline
\bf 0            &  0      &  0.06 &  0      &  0.01 &  0        \\
\bf 2.39       &  0.06 &  0      &  0.56 &  0      &  0.24   \\
\bf 2.41       &  0      &  0.56 &  0      &  0.24 &  0        \\
\bf 5.46       &  0.01 &  0      &  0.24 &  0      &  0.37   \\
\bf 5.49       &  0      &  0.24 &  0      &  0.37      &  0   \\
\hline
\end{tabular}
%
\begin{tabular}{|ccccc|}
\multicolumn{5}{l}{$\bf g^m_{1jk}$} \\ [0.4em]
\hline
\rule{0mm}{4mm}
\bf 0  & \bf 2.39  & \bf 2.41  & \bf 5.46  & \bf 5.49  \\ [0.2em]
\hline
\hline
0      &  0      &  0      &  0      &  0        \\
0      &  0      &  0.22 &  0      &  0.17   \\
0      &  0.22 &  0      &  0.17 &  0        \\
0      &  0      &  0.17 &  0      &  0.31   \\
0      &  0.17 &  0      &  0.31 &  0        \\
\hline
\end{tabular}
\end{center}
\caption{Radion (left) and KK-radion (right) couplings to KK fermion
pairs [see Eq.~(\ref{Lrffbulk})] in the ``strong warping scenario'',
for $c = -0.2$.  We also show the KK-radion-gauge boson couplings [see
Eq.~(\ref{VVrpgijk})].  The lowest lying KK fermion masses, in units
of $\tilde{k}_{\rm eff} = k_{\rm eff}\,e^{-A(L)}$, are $x^{f}_{0} =
0$, $x^{f}_{1} \approx 1.15$, $x^{f}_{2} \approx 3.38$, $x^{f}_{3}
\approx 4.32$, $x^{f}_{4} \approx 6.50$ and $x^{f}_{5} \approx 7.43$,
while the gauge boson ones are $x^{V}_{0} = 0$, $x^{V}_{1} \approx
2.39$, $x^{V}_{2} \approx 2.41$, $x^{V}_{3} \approx 5.46$ and
$x^{V}_{4} \approx 5.49$.  The states with mass $x_{n}$ with even
(odd) $n$ are KK-parity even (odd).  In general, EWSB will lift
$x_{0}$ to a non-vanishing value.  The radion and KK-radion masses are
$m_{0} \approx m_{1} \approx 0.22\,\tilde{k}_{\rm eff}$.  The
dimensionless values in this table were computed assuming $\tilde{k} =
2~{\rm TeV}$ (which corresponds to $\tilde{k}_{\rm eff} \approx
1.2~{\rm TeV}$), but are rather insensitive to this choice, as long as
$\tilde{k} \sim {\cal O}({\rm TeV})$.}
\label{radionCoupStrongWarp}
\end{table}%

\begin{table}[t]
\vspace*{-2cm}
\begin{center}
\begin{tabular}{|c||cccccc|}
\multicolumn{7}{l}{$\bf g^{LL}_{0jk}$} \\ [0.4em]
\hline
\rule{0mm}{4mm}
$x^{f}_{n}$  & \bf 0  & \bf 0.97  & \bf 2.96  & \bf 3.78  & \bf 5.63  & \bf 6.51  \\ [0.2em]
\hline
\hline
\bf 0            &  0      &  0      &  0       &  0      &  0      & 0       \\
\bf 0.97       &  0      &  0.06 &  0      &  0.11 &  0      & 0.03  \\
\bf 2.96       &  0      &  0      &  0.24 &  0      &  0.17 & 0       \\
\bf 3.78       &  0      &  0.11 &  0      &  0.27 &  0      & 0.18  \\
\bf 5.63       &  0      &  0      &  0.17 &  0      &  0.27 & 0       \\
\bf 6.51       &  0      &  0.03 &  0      &  0.18 &  0      & 0.28  \\
\hline
\end{tabular}
%
\begin{tabular}{|cccccc|}
\multicolumn{6}{l}{$\bf g^{LL}_{1jk}$} \\ [0.4em]
\hline
\rule{0mm}{4mm}
\bf 0  & \bf 0.97  & \bf 2.96  & \bf 3.78  & \bf 5.63  & \bf 6.51  \\ [0.2em]
\hline
\hline
0      &  0      &  0      &  0      &  0      & 0       \\
0      &  0      &  0.11 &  0      &  0.05 & 0       \\
0      &  0.11 &  0      &  0.25 &  0      & 0.13  \\
0      &  0      &  0.25 &  0      &  0.22 & 0       \\
0      &  0.05 &  0      &  0.22 &  0      & 0.27  \\
0      &  0      &  0.13 &  0      &  0.27 & 0       \\
\hline
\end{tabular}
\end{center}
\begin{center}
\begin{tabular}{|c||cccccc|}
\multicolumn{7}{l}{$\bf g^{RR}_{0jk}$} \\ [0.4em]
\hline
\rule{0mm}{4mm}
$x^{f}_{n}$  & \bf 0  & \bf 0.97  & \bf 2.96  & \bf 3.78  & \bf 5.63  & \bf 6.51  \\ [0.2em]
\hline
\hline
\bf 0            &  0.38 &  0      &  0.28 &  0      &  0.10 & 0       \\
\bf 0.97       &  0      &  0.48 &  0      &  0.26 &  0      & 0.09  \\
\bf 2.96       &  0.28 &  0      &  0.32 &  0      &  0.21 & 0       \\
\bf 3.78       &  0      &  0.26 &  0      &  0.31 &  0      & 0.21  \\
\bf 5.63       &  0.10 &  0      &  0.21 &  0      &  0.29 & 0       \\
\bf 6.51       &  0      &  0.09 &  0      &  0.21 &  0      & 0.30  \\
\hline
\end{tabular}
%
\begin{tabular}{|cccccc|}
\multicolumn{6}{l}{$\bf g^{RR}_{1jk}$} \\ [0.4em]
\hline
\rule{0mm}{4mm}
\bf 0  & \bf 0.97  & \bf 2.96  & \bf 3.78  & \bf 5.63  & \bf 6.51  \\ [0.2em]
\hline
\hline
0      &  0.42 &  0      &  0.22 &  0      & 0.07  \\
0.42 &  0      &  0.33 &  0      &  0.13 & 0       \\
0      &  0.33 &  0      &  0.30 &  0      & 0.16  \\
0.22 &  0      &  0.30 &  0      &  0.25 & 0       \\
0      &  0.13 &  0      &  0.25 &  0      & 0.28  \\
0.07 &  0      &  0.16 &  0      &  0.28 & 0       \\
\hline
\end{tabular}
\end{center}
\begin{center}
\begin{tabular}{|c||cccccc|}
\multicolumn{7}{l}{$\bf m^{LR}_{0jk}/\tilde{k}_{\rm eff}$} \\ [0.4em]
\hline
\rule{0mm}{4mm}
$x^{f}_{n}$  & \bf 0  & \bf 0.97  & \bf 2.96  & \bf 3.78  & \bf 5.63  & \bf 6.51  \\ [0.2em]
\hline
\hline
\bf 0            &  0      &  0      &  0      &  0      &  0      & 0       \\
\bf 0.97       &  0      &  1.06 &  0      &  1.31 &  0      & 0.58  \\
\bf 2.96       &  1.70 &  0      &  3.34 &  0      &  3.17 & 0       \\
\bf 3.78       &  0      &  2.24 &  0      &  4.40 &  0      & 3.87  \\
\bf 5.63       &  1.14 &  0      &  3.43 &  0      &  6.38 & 0       \\
\bf 6.51       &  0      &  1.28 &  0      &  4.09 &  0      & 7.59  \\
\hline
\end{tabular}
%
\begin{tabular}{|cccccc|}
\multicolumn{6}{l}{$\bf m^{LR}_{1jk}/\tilde{k}_{\rm eff}$} \\ [0.4em]
\hline
\rule{0mm}{4mm}
\bf 0  & \bf 0.97  & \bf 2.96  & \bf 3.78  & \bf 5.63  & \bf 6.51 \\ [0.2em]
\hline
\hline
0      &  0      &  0      &  0      &  0      & 0       \\
0.84 &  0      &  1.31 &  0      &  0.82 & 0       \\
0      &  2.22 &  0      &  3.68 &  0      & 2.59  \\
1.67 &  0      &  3.78 &  0      &  4.40 & 0       \\
0      &  1.59 &  0      &  4.56 &  0      & 6.66  \\
0.93 &  0      &  2.91 &  0      &  6.71 & 0       \\
\hline
\end{tabular}
\end{center}
\begin{center}
\begin{tabular}{|c||ccccc|}
\multicolumn{6}{l}{$\bf g_{1jk}$} \\ [0.4em]
\hline
\rule{0mm}{4mm}
$x^{V}_{n}$  & \bf 0  & \bf 2.19  & \bf 2.41  & \bf 4.86  & \bf 5.17  \\ [0.2em]
\hline
\hline
\bf 0            &  0      &  0.16 &  0      &  0.04 &  0        \\
\bf 2.19       &  0.16 &  0      &  0.51 &  0      &  0.22   \\
\bf 2.41       &  0      &  0.51 &  0      &  0.26 &  0        \\
\bf 4.86       &  0.04 &  0      &  0.26 &  0      &  0.33        \\
\bf 5.17       &  0      &  0.22 &  0      &  0.33 &  0   \\
\hline
\end{tabular}
%
\begin{tabular}{|ccccc|}
\multicolumn{5}{l}{$\bf g^m_{1jk}$} \\ [0.4em]
\hline
\rule{0mm}{4mm}
\bf 0  & \bf 2.19  & \bf 2.41  & \bf 4.86  & \bf 5.17  \\ [0.2em]
\hline
\hline
0      &  0      &  0      &  0      &  0        \\
0      &  0      &  0.19 &  0      &  0.15   \\
0      &  0.19 &  0      &  0.18 &  0        \\
0      &  0      &  0.18 &  0      &  0.27   \\
0      &  0.15 &  0      &  0.27 &  0        \\
\hline
\end{tabular}
\end{center}
\caption{Radion (left) and KK-radion (right) couplings to KK fermion
pairs [see Eq.~(\ref{Lrffbulk})] in the ``small warping scenario'',
for $c = -0.2$.  We also show the KK-radion-gauge boson couplings [see
Eq.~(\ref{VVrpgijk})].  The lowest lying KK fermion masses, in units
of $\tilde{k}_{\rm eff} = k_{\rm eff}\,e^{-A(L)}$, are $x^{f}_{0} =
0$, $x^{f}_{1} \approx 0.97$, $x^{f}_{2} \approx 2.96$, $x^{f}_{3}
\approx 3.78$, $x^{f}_{4} \approx 5.63$ and $x^{f}_{5} \approx 6.51$,
while the gauge boson ones are $x^{V}_{0} = 0$, $x^{V}_{1} \approx
2.19$, $x^{V}_{2} \approx 2.41$, $x^{V}_{3} \approx 4.86$ and
$x^{V}_{4} \approx 5.17$.  The states with mass $x_{n}$ with even
(odd) $n$ are KK-parity even (odd).  In general, EWSB will lift
$x_{0}$ to a non-vanishing value.  The radion and KK-radion masses are
$m_{0} \approx m_{1} \approx 0.66\,\tilde{k}_{\rm eff}$.  The
dimensionless values in this table were computed assuming $\tilde{k} =
2~{\rm TeV}$ (which corresponds to $\tilde{k}_{\rm eff} \approx
500~{\rm GeV}$), but are not very sensitive to this choice, as long as
$\tilde{k} \sim {\cal O}({\rm TeV})$.}
\label{radionCoupSmallWarp}
\end{table}%

\subsection{KK-Radion Couplings to Gauge Bosons and the Higgs Fields}
\label{sec:RAAcouplings}

We also summarize the Feynman rules for one or two radion KK modes and
two gauge KK modes.  We focus on the interactions with $V^n_{\mu}$,
which are sufficient for tree-level calculations (where one can work
in unitary gauge with $V^{n}_{5} = 0$).  The $F_{\mu\nu}F^{\mu\nu}$
terms in Eq.~(\ref{Sgauge}) give
\bea
-\frac{1}{4} \frac{1}{2L} \sum_{j,k=0}^{\infty} [ 1+2F(x,y) ]
f^{j}_{V} f^{k}_{V} F^{j}_{\mu\nu}(x) F^{\mu\nu k}(x)~,
\label{RadionGaugemunu}
\eea
(with indices raised by the Minkowski metric), while the $F_{\mu5}
F^{\mu5}$ terms give
\bea
\frac{1}{2} \frac{1}{2L} \sum_{j,k=0}^{\infty}
e^{-2[A(y)+F(x,y)]} [ 1+2F(x,y) ]^{-1} \,
\partial_{y} f^{j}_{V} \partial_{y}f^{k}_{V} \, V^{j}_{\mu}(x) V^{\mu k}(x)~,
\label{RadionGauge5mu}
\eea
where we used the parameterization of the radion modes given in
Eq.~(\ref{MetricFluctuations}), as well as the gauge KK
decomposition, Eq.~(\ref{GaugeKK}).  Using also the radion KK
decomposition of Eq.~(\ref{RadionKKDecomposition}), with $r_{i}(x)
\equiv \Lambda_{r} \tilde{r}_{i}(x)$, the Feynman rule for the
vertex involving a single radion and two gauge fields reads
\bea
\put(-30,-30){
\resizebox{3.2cm}{!}{
\includegraphics{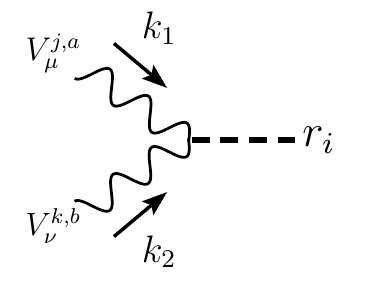}}
}
\hspace*{2cm}
&=&
\frac{2i}{\Lambda_r} \delta^{ab} \left\{g_{ijk} \left[ k_{1} \! \cdot \! k_{2} \, \eta_{\mu\nu} - k_{1\mu} k_{2\nu} \right]
-2 g^m_{ijk} m_{j}m_{k} \, \eta_{\mu\nu} \right\}~,
\label{VVrp}
\eea
where
\bea
g_{ijk}=\frac{1}{2L}\int^{L}_{-L} \! dy \, F_{i}f^j_{V}f^{k}_{V}~,
\hspace{1cm}
g^m_{ijk}=\frac{1}{2L}\int^{L}_{-L} \! dy \, e^{-2A} F_{i}f^j_{5}f^{k}_{5}~,
\label{VVrpgijk}
\eea
and we restored the gauge indices, $a,b$.  Here we used
$f_5^{j}=\partial_y f^j_{V}/m_j$, since it is simpler to find
$f_5^{j}(y)$ numerically.~\footnote{More concretely, in order to find
the gauge KK spectrum, $m_{j}$, we apply the ``shooting method''
described at the end of Subsection~\ref{sec:KKFermion} to
$\tilde{f}^{j}_{5} \equiv e^{-2A} f^{j}_{5}$, which --from
Eq.~(\ref{GaugeEOM})-- obeys $\partial^{2}_{y}\tilde{f}_{5}^{j} +
m_{j}^{2} e^{2A} \tilde{f}^{j}_{5} = 0$ and $\left.  \tilde{f}^{j}_{5}
\right|_{\pm L} = 0$.} Note that Eqs.~(\ref{VVrpgijk}) with $F_{i} \to
1$ are precisely the orthonormality relations for the $f^{j}_{V}$ and
$f^{j}_{5}$.  For completeness, we also quote the interactions of two
radions and two gauge bosons, which arise from
Eq.~(\ref{RadionGauge5mu}), although these do not involve the gauge
zero-modes:
\bea
\put(-30,-30){
\resizebox{3.2cm}{!}{
\includegraphics{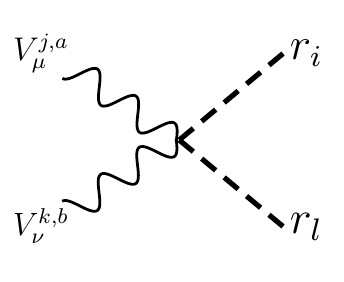}}
}
\hspace*{2cm}
&=&
\frac{20i}{\Lambda_r^2} \delta^{ab}  g^m_{iljk} m_{j}m_{k} \, \eta_{\mu\nu}~,
\label{VVrprp}
\eea
where $g^m_{iljk}$ is defined as $g^m_{ijk}$ in Eq.~(\ref{VVrpgijk}),
but with two radion wavefunctions, $F_{i} F_{l}$.

Finally, we give the Feynman rules for the Higgs interactions with one
and two radion modes.  Here we assume, for simplicity, that the Higgs
fields are IR localized, so that the radion couplings enter through
the induced metric.  We also neglect the possible mixing between
Higgses and radion modes.  Referring to the discussion of kinetic and
potential Higgs terms of Subsection~\ref{sec:EWSB}, one must only
remember that the radion modes enter through a multiplicative factor
$e^{-2F(x,y)}$ when coupled to scalar kinetic terms, and through
$e^{-4F(x,y)}$ when coupled to potential terms.  Using the radion KK
decomposition, Eq.~(\ref{RadionKKDecomposition}), and in terms of the
Higgs mass eigenstates defined in Eq.~(\ref{HiggsDoublets}), the
Feynman rules involving a single radion mode are
\bea r_{i} \phi \phi  &\longrightarrow& i \left[1 +
(-1)^{i}\right] \frac{F_{i}(L)}{\Lambda_{r}} \left[ k_{1} \! \cdot
\! k_{2} + 2 m^{2}_{\phi} \right]~,
\\[0.5em]
r_{i} \, h_{+} h_{-}  &\longrightarrow& i \left[1 -
(-1)^{i}\right] \frac{F_{i}(L)}{\Lambda_{r}} \left[ k_{1} \! \cdot
\! k_{2} + 4 m^{2}_{h_{+}} \right]~,
\\[0.5em]
r_{i} \, (a,G^{0})/(H^{\pm} G^{\mp})  &\longrightarrow& i \left[1
- (-1)^{i}\right] \frac{F_{i}(L)}{\Lambda_{r}} \, k_{1} \! \cdot
\! k_{2}~, \eea
where $\phi = h_{+}, h_{-}, a, H^{\pm}, G^{0},G^{\pm}$ (and
$m_{G^{0},G^{\pm}}^{2}=0$).  We also denoted the (incoming) Higgs
momenta by $k_{1}$ and $k_{2}$, and $F_{i}(y)$ is the $r_{i}$
wavefunction.  The Feynman rules involving two radion modes are
\bea r_{i} r_{j} \phi \phi  &\longrightarrow& -2i \left[1 +
(-1)^{i+j}\right] \frac{F_{i}(L) F_{j}(L)}{\Lambda_{r}^{2}} \left[
k_{1} \! \cdot \! k_{2} + 4 m^{2}_{\phi} \right]~,
\\[0.5em]
r_{i} r_{j} h_{+} h_{-} &\longrightarrow& -2i \left[1 -
(-1)^{i+j}\right] \frac{F_{i}(L) F_{j}(L)}{\Lambda_{r}^{2}} \left[
k_{1} \! \cdot \! k_{2} + 8 m^{2}_{h_{+}} \right]~,
\\[0.5em]
r_{i} r_{j} \, (a,G^{0})/(H^{\pm} G^{\mp}) &\longrightarrow& -2i
\left[1 - (-1)^{i+j}\right] \frac{F_{i}(L)
F_{j}(L)}{\Lambda_{r}^{2}} \, k_{1} \! \cdot \! k_{2}~.
\eea

\section{Summary and Conclusions}
\label{sec:conclusions}

In this work we have introduced a 5-dimensional scenario where the
dynamics that stabilizes the size of the fifth dimension: i) induces
the non-trivial warping that allows understanding the weakness of the
gravitational interactions compared to the weak interactions ii) leads
to a metric background that implies the existence of a KK-parity
symmetry, thus predicting a weak scale stable particle (the LKP), iii)
leads to fermion localization via Yukawa interactions involving the
stabilizing field, and iv) the Higgs sector is described by a THDM
with an ``inert'' Higgs doublet ($\tan\beta = \infty$).  It is argued
that generically the LKP is the first excitation associated with the
radion field.  This KK-radion is expected to be highly degenerate with
the radion, and its interactions are controlled by the same decay
constant that controls the radion interactions.  This allows to infer
properties of the KK-radion from collider signatures of the radion,
and help identify the collider KK-radion signatures.  In particular,
it may be possible to probe whether the LKP accounts for the DM
content of the universe via collider measurements.  Although we have
not explored this question in this paper (we present the analysis in
the companion paper~\cite{Medina:2011qc}), we have presented the
technical ingredients necessary to study the phenomenology of such
scenarios.  In particular, we provided simple analytical expressions
for the masses and wavefunctions of the lowest lying KK states, which
can be summarized as follows:
\begin{itemize}

\item The KK scale is set by the warped down IR curvature scale,
$\tilde{k}_{\rm eff} = k_{\rm eff} \, e^{-A(L)} = A'(L) \, e^{-A(L)}$,
where $2L$ is the proper size of the extra dimension, while $A(y)$
depends on the stabilization mechanism (and needs not be linear in
$y$, as is the case of AdS$_{5}$).

\item The radion (KK-radion) is strongly and symmetrically
(antisymmetrically) localized near the IR boundaries, according to
$F_0(y) \approx e^{2[A(y)-A(L)]}$ for $y \in [0,L]$.  Their masses are
exponentially degenerate at tree level, and given approximately by
$\frac{2}{\sqrt{k_{\rm eff} L}} \, \tilde{k}_{\rm eff}$ in the case
that the dynamical ``UV brane'' is fat [see Eq.~(\ref{mRadionExact})
for a completely general expression].

\item The zero-mode fermions have wavefunctions that behave like
$e^{-\left( c - \frac{1}{2} \right) A(y) }$ in the ``fat UV brane''
limit [see Eqs.~(\ref{zeromodeSimple}) and (\ref{fermionN1N0}), and
Eq.~(\ref{zeromodes}) for the exact expression in a general 4D Lorentz
invariant background].  The effective $c$-parameter allows a simple
characterization of the localization properties.  In the ``fat UV
brane'' limit, UV localized wavefunctions ($c \gtrsim 1/2$) are
Gaussian-like, while IR localized wavefunctions ($c \lesssim 1/2$) are
approximately exponential.

\item The lightest KK fermion excitations that are localized towards
the IR boundaries have a mass approximately given by $m_1 \approx
\sqrt{2(1 + 2c)} \, \tilde{k}_{\rm eff}$ when $-1/2 \lesssim c \lesssim
1/2$.  Their LH and RH wavefunctions were given in Eqs.~(\ref{f1L})
and (\ref{f1R}).

\end{itemize}

We also provided sample KK spectra and couplings between
radion/KK-radion and fermion/gauge modes in
Tables~\ref{radionCoupStrongWarp} and \ref{radionCoupSmallWarp}, and
briefly discussed the fermion couplings to the Higgs, as well as the
KK-radion/Higgs interactions.  These ingredients are relevant when
studying the phenomenology of the present scenario.  We discussed the
possibility that the curvature scale ranges from order the fundamental
Planck scale down to much smaller values.  The former case is similar
in spirit to the RS proposal, while the latter has some features in
common with Universal Extra Dimensions, but allows a consistent
description of the gravitational interactions and, via the AdS/CFT
correspondence may be a valid description up to a scale exponentially
larger than the weak scale (but well below the 4D Planck mass).  This
latter scenario may allow for a lighter spectrum of KK states than is
expected in RS models.  The EW and flavor constraints deserve a more
detailed study, as does the associated collider phenomenology of the
$Z_{2}$-symmetric warped scenarios.

\section*{Acknowledgments}

We would like to thank Thomas Flacke and especially Arjun Menon for
collaboration at early stages of this work.  We also thank Hsin-Chia
Cheng for discussions.  A.M. is supported by the US department of
Energy under contract DE-FG02-91ER406746.  E.P. is supported by DOE
grant DE-FG02-92ER40699.

\end{document}